\documentclass[a4paper,fleqn,usenatbib]{mnras}

\usepackage[T1]{fontenc}
\usepackage{ae,aecompl}
\usepackage{graphicx}	
\usepackage{amsmath}	
\usepackage{amssymb}	

\usepackage{slashbox}
\usepackage{multirow}
\usepackage{txfonts}
\usepackage{lscape}

\title[The mid-infrared interferometric catalogue]{A catalogue of stellar diameters and fluxes for mid-infrared interferometry\thanks{The catalogue is available in electronic form at the OCA/MATISSE webpage via
https://matisse.oca.eu/foswiki}}

\author[P. Cruzal\`{e}bes et al.]{
P. Cruzal\`{e}bes,$^{1}$\thanks{E-mail: pierre.cruzalebes@oca.eu}
R.G. Petrov,$^{1}$
S. Robbe-Dubois,$^{1}$
J. Varga,$^{2,3}$
L. Burtscher,$^{2}$
F. Allouche,$^{1}$
\newauthor{
P. Berio,$^{1}$
K.-H. Hofmann,$^{4}$
J. Hron,$^{5}$
W. Jaffe,$^{2}$
S. Lagarde,$^{1}$
B. Lopez,$^{1}$
A. Matter,$^{1}$}
\newauthor{
A. Meilland,$^{1}$
K. Meisenheimer,$^{6}$
 F. Millour,$^{1}$
 and D. Schertl$^{4}$}\\
\\
$^{1}$Universit\'{e} C\^{o}te d'Azur, Observatoire de la C\^{o}te d'Azur, CNRS, Laboratoire Lagrange, Parc Valrose, B\^{a}t. H. Fizeau, 06108 Nice, France\\
$^{2}$Leiden Observatory, Leiden University, Niels Bohrweg 2, 2333 CA Leiden, The Netherlands\\
$^{3}$Konkoly Observatory, Research Centre for Astronomy and Earth Sciences, Hungarian Academy of Sciences, Konkoly Thege Mikl\'{o}s \'{u}t 15-17, 1121  Budapest, Hungary\\
$^{4}$Max-Planck-Institut f\"{u}r Radioastronomie, Auf dem H\"{u}gel 69, 53121 Bonn, Germany\\
$^{5}$Department of Astrophysics, University of Vienna, T\"{u}rkenschanzstrasse 17, 1180 Vienna, Austria\\
$^{6}$Max-Planck-Institut f\"{u}r Astronomie, K\"{o}nigstuhl 17, 69117 Heidelberg, Germany
}

\date{Accepted XXX. Received YYY; in original form ZZZ}

\pubyear{2019}

\begin{document}
\label{firstpage}
\pagerange{\pageref{firstpage}--\pageref{lastpage}}
\maketitle

\begin{abstract}
We present the Mid-infrared stellar Diameters and Fluxes compilation Catalogue (MDFC) dedicated to long-baseline interferometry at mid-infrared wavelengths (3-13~$\mu$m). It gathers data for half a million stars, i.e. nearly all the stars of the \textit{Hipparcos}-Tycho catalogue whose spectral type is reported in the SIMBAD database. We cross-match 26 databases to provide basic information, binarity elements, angular diameter, magnitude and flux in the near and mid-infrared, as well as flags that allow us to identify the potential calibrators. The catalogue covers the entire sky with 465\,857 stars, mainly dwarfs and giants from B to M spectral types closer than 18~kpc. The smallest reported values reach 0.16~$\mu$Jy in $L$ and 0.1~$\mu$Jy in $N$ for the flux, and 2~microarcsec for the angular diameter. We build 4 lists of  calibrator candidates for the $L$- and $N$-bands suitable with the Very Large Telescope Interferometer (VLTI) sub- and main arrays using the MATISSE instrument. We identify 1\,621 candidates for $L$ and 44 candidates for $N$ with the Auxiliary Telescopes (ATs), 375 candidates for both bands with the ATs, and 259 candidates for both bands with the Unit Telescopes (UTs).  Predominantly cool giants, these sources are small and bright enough to belong to the primary lists of calibrator candidates. In the near future, we plan to measure their angular diameter with 1\% accuracy. 
\end{abstract}




%
\section{Introduction}

Modern long-baseline
    interferometers, such as the ESO-VLTI \citep{paresce96} or the CHARA \citep{brummelaar05} arrays,
 combine the beams of many telescopes (4 for the VLTI, 6 for CHARA) to achieve high angular resolution observations using hectometric baselines. They measure the spectral variation of the correlated flux,
    visibility, closure phase, differential visibility and  phase in the specific spectral bands.

MIDI, the MID-infrared Interferometric instrument \citep{leinert03}, was the first scientific instrument available at the VLTI covering the photometric $N$-band (from 8 to 13~$\mu$m). Decommissioned in late 2015, it is now replaced by the second-generation instrument MATISSE, the Multi AperTure mid-Infrared SpectroScopic Experiment \citep{lopez14,allouche16,robbe18} which operates simultaneously in the 3 photometric bands: $L$ (from 2.8 to 4.2~$\mu$m), $M$ (from 4.5 to 5~$\mu$m), and $N$ (from 8 to 13~$\mu$m).

Reference targets with well-known angular diameters and hence absolute visibilities are essential to calibrate interferometric observables. Up to now the the JMMC Stellar Diameters Catalogue \citep[JSDC, Cat. II/346;][]{bourges17} is the largest database of computed angular diameters  in the literature, for nearly half a million stars. It is a basic input of our work. It reports angular diameter estimates of the limb-darkened disk (LDD) and uniform disk (UD) from the $B$ to $N$  photometric bands. It also reports visible magnitudes from the Tycho-2 Catalogue \citep[Cat. II/259;][]{hog00}, near-infrared (NIR) magnitudes from the 2MASS All-Sky Catalogue of Point Sources \citep[Cat. II/246;][]{cutri03}, and mid-infrared (MIR) magnitudes from  the WISE All-Sky Data Release catalogue \citep[Cat. II/311;][]{wright10}. In the framework of the scientific exploitation of the VLTI/MATISSE instrument, we need the most reliable estimates of both the angular diameter and the flux density for MIR wavelengths. As we point out in Section~\ref{section:MIRflux}, a significant number of sources are reported in the WISE catalogue as having spurious MIR flux values \citep[see e.g.][]{cutri12}.

The need of reliable flux values leads us to propose a new all-sky catalogue, the Mid-infrared stellar Diameters and Fluxes compilation
    Catalogue (MDFC), that gathers data from various existing literary sources in order to report  in the same table: 
\begin{itemize}
\item angular diameter estimates and measurements; 
\item flux density measurements and estimates at MIR wavelengths;
\item flags identifying the potential interferometric calibrators suitable for the MIR spectral domain.
\end{itemize}
The main advantages of our catalogue over the JSDC are to supply reliable flux values in the MIR and to report a calibrator selection criterion suitable for this spectral domain.

The paper is divided in 5 main sections. Section~\ref{section:build} presents the successive steps followed to build the MDFC.  Section~\ref{section:IRflag} describes the new flag that allows the identification of the sources that show infrared extended structures (excess, extent, variability).
 Section~\ref{section:results} presents the content of the MDFC in terms of sky-coverage and distributions in: spectral type, distance, binaries, angular diameter, MIR flux, and calibrators. Section \ref{section:calib} gives clues for selecting calibrators for observations with the VLTI/MATISSE instrument. Section~\ref{section:appli} shows an example of application of the catalogue with the building of the primary lists of calibrators for VLTI/MATISSE.

\section{Building the catalogue} \label{section:build}

To build the MDFC:
\begin{itemize}
\item we use as input the JDSC catalog which reports angular diameter estimates with their uncertainties;
\item we complete the diameter data with those of 3 other diameter catalogs providing measurements and other estimates; 
\item we include the basic data (including astrometric binarity);
\item we compile MIR flux data reported in 12 other photometric catalogues;
\item we add a new flag that identifies the stars showing MIR features (excess, extent, variability).
\end{itemize}

\subsection{Input catalogues for the angular diameter}
   Since the JSDC is the most complete and up-to-date catalog used for visible/infrared interferometric calibration, we use it as primary input to build the MDFC. The JSDC contains 465\,877 entries (among which we have identified and deleted 272 duplicates) and reports estimates of  LDD and UD angular diameters from the $B$ to $N$ spectrophotometric bands, providing a flag for each star indicating a degree of confidence in choosing it as 
    a calibrator for optical long-baseline interferometry (OLBI). We complete the diameter data with the values reported in the 3 following smaller catalogs, also widely used for interferometric calibration:
\begin{enumerate}
   \item the JMDC, the JMMC Measured Stellar Diameters Catalogue \citep[Cat. II/345;][]{duvert16}, that reports 1554  direct measurements of UD and LDD diameters for 566 different stars, made with "direct" techniques (optical interferometry, intensity interferometry, and lunar occultations) from the visible to the MIR wavelengths; 
   \item the VLTI/MIDI list of calibrator candidates, which reports 403 estimates of LDD diameter with fitted effective temperatures obtained from \citet{verhoelst05};
   \item the Cohen's list of spectrophotometric standards, which reports 422 estimates of LDD diameter derived from calibrated spectral templates \citep[][Table 4]{cohen99}.
 \end{enumerate}  
  The aggregating of these 3 small catalogues with the JSDC leads to a total of 465\,857 stars that have at least one angular diameter estimate or measurement reported at visible and/or infrared wavelengths. Note that if many diameter values are available for a given star, we suggest to favour the value given by the JSDC primary catalog. Large discrepancies between angular diameter estimates for a given source may indicate that this source is a bad calibrator. 
  The final catalog includes nearly all the stars of the
    \textit{Hipparcos}-Tycho catalogue \citep[Cat. I/239;][]{esa97} whose spectral type is reported in the SIMBAD database \citep{wenger00}. All entries of the catalogue have $H$ and $K$ magnitudes reported in 2MASS.

\subsection{Reporting basic, fundamental, and binarity data} 
    Spectral type and equatorial coordinates are taken from the SIMBAD database. Binarity observational data (astrometric) is reported from the Washington Double Star Catalogue \citep[Cat. B/wds;][]{mason01}. \\
We report the effective temperature and the stellar physical radius from the \textit{Gaia} DR2 catalogue \citep[Cat. I/345/gaia2;][]{gaia18}, and the distance from the complement of the \textit{Gaia} DR2 catalogue \citep[Cat. I/347/gaia2dis;][]{bailerjones18}. From the radius to distance ratio we report another estimate of the LDD diameter\footnote{The values of effective temperature and radius reported in the \textit{Gaia} DR2 catalogue were determined only from the three broad-band photometric measurements with \textit{Gaia}. The strong degeneracy between $T_\mathrm{eff}$ and extinction/reddening when using the broad-band photometry necessitates strong assumptions in order to estimate their values \citep[see e.g.][]{casagrande18}. One should thus be very careful in using these astrophysical parameters and refer to the papers and online documentation for guidance}.

\subsection{Reporting and merging MIR flux data}\label{section:MIRflux}

\begin{figure}
   \centering
  \includegraphics[trim = 5cm 0cm 3cm 0cm, clip, width=\hsize]{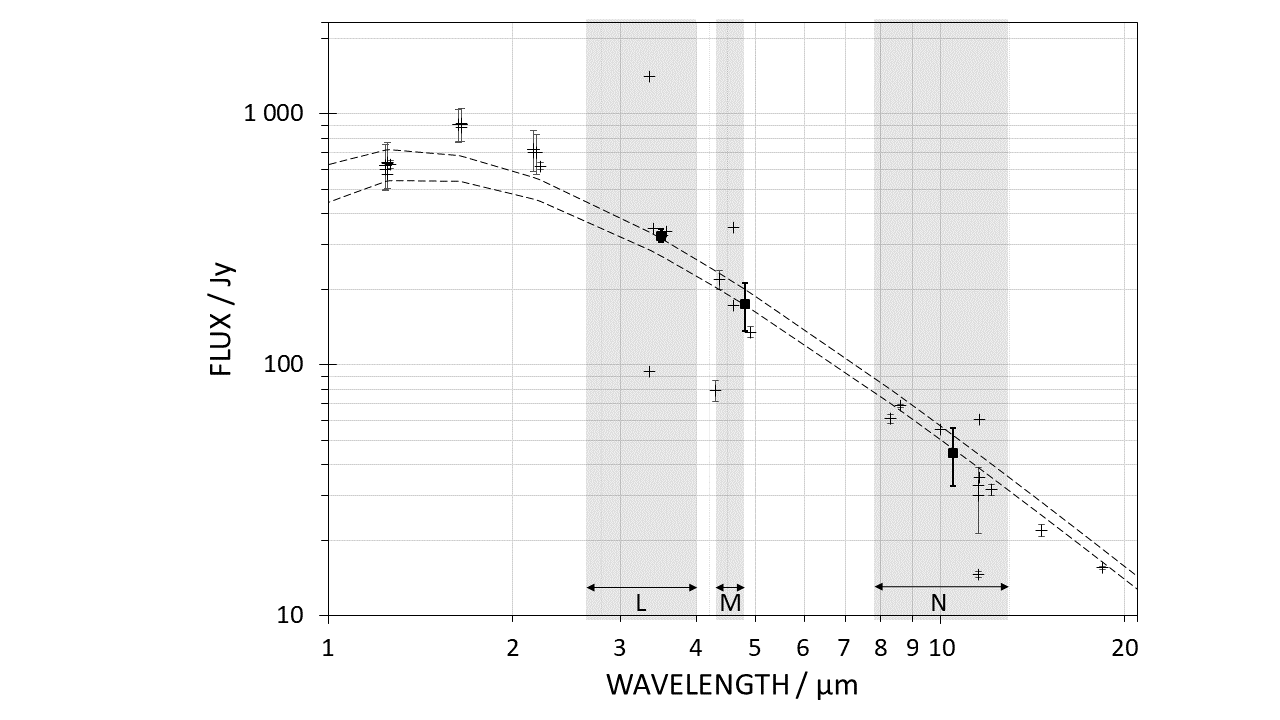}
      \caption{Flux measurements between 1 and 20~$\mu$m reported in the VizieR Photometry viewer for the M3II star 72~Leo  in a radius of 5\arcsec around its central position. The spectral domains covered by the VLTI/MATISSE are shaded in grey. The 3 black squares are the median values reported in our catalogue with their statistical dispersion. The upper and lower dashed curves are the SEDs given by the backbody model with the MIDI and \textit{Gaia} DR2 parameters respectively. }
         \label{fig:viewer}
   \end{figure}

    Beside basic and angular diameter data, our catalogue
    also reports flux data in the NIR and MIR. The $J$, $H$, and $K$ magnitudes
    are reported in the 2MASS Catalogue. For a substantial number of stars (mainly IR-bright stars), we notice that the flux measurements in the MIR photometric bands may vary significantly from one catalogue to the other. In the example of the M3II star 72~Leo shown in Fig.~\ref{fig:viewer}, we note some spurious flux values reported in the MIR bands. The wide dispersion of the measurements observed for these sources is greater than the spread in flux caused by the width of the spectral band. Under the blackbody model assumption with $T_\mathrm{eff}$~=~4000~K, this spread in flux does not exceed 43\% for the $L$-band (2.8-4.2~$\mu$m), 15\% for the $M$-band (4.5-5~$\mu$m), and 58\% for the $N$-band (8-13~$\mu$m).

For each star, we compile the flux values reported in various catalogues. Our goal is to report an approximate but reliable final value of the flux density for each MIR photometric band, without any use of modeling the in-band flux based on the measured photometry. If more than 2 flux values are reported for each MIR band, we compute the median value ($F$), less sensitive to spurious values than the sample mean. The reliability of the final flux value is estimated by the median absolute deviation from the median ($MAD$), which gives a robust estimate of the statistical scatter. If only 2 flux values are reported, we compute the mean value and the range, i.e. the difference between the maximum and the mininimum values. If one of those 2 values is irrelevant, the large value of the range is an indication of the irrelevance of the final flux value. If only one flux value is reported, no associated dispersion is reported.

The flux density measurements are reported in: 
\begin{itemize}
    \item the AllWISE survey \citep[Cat. II/328;][747 million sources covering 99.86\% of the entire sky]{wright10} for the WISE/W1, W2, and W3 filters, with flux sentitivities better than 0.08, 0.11, and 1~mJy respectively. Since the original WISE 
    catalogue (563 million sources) may have better photometric information for objects brighter than the saturation limit for the W1 and W2 filters, the WISE flux measurements supersede the AllWISE measurements for magnitudes [W1]~<~8 and [W2]~<~7  \citep[see][]{cutri13}; 
    \item the GLIMPSE Source Catalogue
      \citep[Cat. II/293;][104 million sources covering approximately 220 square degrees for $|l|$~$\le$~65\degr-- galactic longitude -- and $|b|$~$\le$~1\degr -- galactic latitude]{benjamin03} for the 3.6-, 4.5-, and 8-$\mu$m \textit{Spitzer}-IRAC bands, with flux sensitivities better than 0.6, 0.4, and 10~mJy respectively;
   \item the AKARI/IRC All-Sky Survey
      Point Source Catalogue \citep[Cat. II/297;][870\,973 sources covering more than 90\% of the entire sky]{ishihara10}  for the AKARI/S9W filter with flux sentitivity better than 20~mJy;
    \item the MSX6C Infrared Point Source
      Catalogue \citep[Cat. V/114;][431\,711 sources for $|b|$~$\le$~6\degr and 10\,168 sources for $|b|$~>~6\degr, covering a total of 10\% of the entire sky]{egan03} for the A-, B1-, B2-, and C-MSX bands, with flux sensitivity reaching  0.1, 10, 6, and 1.1~Jy respectively;
   \item the IRAS PSC/FSC
      Combined Catalogue \citep[Cat. II/338;][345\,162 sources covering the entire sky]{abrahamyan15} for the IRAS/12 filter with flux sentitivity better than 0.25~Jy. The IRAS/12 values are completed with the values of the 
    IRAS Faint Source Catalogue \citep[Cat. II/156A;][173\,044 sources for $|b|$~>~10\degr]{moshir90} with flux sentitivity  better than 0.1~Jy, and those of the Point
      Sources Catalogue \citep[Cat. II/125;][245\,889 sources covering the entire sky]{helou88} with flux sentitivity better than 0.25~Jy;
    \item the COBE DIRBE Point Source
      Catalogue \citep[Cat. J/ApJS/154/673;][11\,788 sources covering 85\% of the entire sky]{smith04} for the F3.5, F4.9, and F12 COBE-DIRBE filters, with flux sensitivities at high galactic latitudes ($|b|$~>~5\degr) better than 60, 50, and 90~Jy respectively;
    \item  the $UBVRIJKLMNH$ Photoelectric
      Catalogue \citep[Cat. II/7A;][compiling reported measurements of 4\,494 sources published up to 1978]{morel78} for the $L$-, $M$-, and $N$-Johnson filters;
   \item the Catalogue of
    10-micron Celestial Objects \citep[Cat. II/53;][compiling reported measurements of 647 sources published between 1964 and 1973]{hall74} for $\lambda$~=~10~$\mu$m.
\end{itemize}
 The conversion of MIR magnitudes to fluxes is done using the zero-magnitude flux values given in Table~\ref{table:filters}, which also gives the angular resolution in each filter used with the different surveys.

%
\begin{table}
\caption{Characteristics of the broad-band filters (-- indicates an undefined value) used to report flux in the bands: $L$ (top), $M$ (middle), and $N$ (bottom).}             
\label{table:filters}      
\centering                          
\begin{tabular}{l c c c c }        
\hline\hline                 
 Catalogue & Filter & Isophotal & Zero-mag.  & Angular  \\    
                &          &  wavel. ($ \mu$m) & flux (Jy) & Resol. \\
\hline                        
   WISE  & W1 & 3.35 & 310 & 6.1\arcsec \\      
   MIDI  & SAAO/L & 3.5  & 294 & 34\arcsec  \\
   JP11  & L & 3.5   & 288 & -- \\
  DIRBE & F3.5 & 3.5   & 282 & 0.7\degr  \\
  GLIMPSE & IRAC3.6 & 3.6   & 281 & 1.7\arcsec \\
\hline
   MSX  & B1 & 4.29  & 195 & 18.3\arcsec \\      
   MSX  & B2 & 4.35    & 189 & 18.3\arcsec \\
   GLIMPSE  & IRAC4.5 & 4.5  & 180 &  1.7\arcsec\\
   WISE & W2 & 4.6  & 172 & 6.4\arcsec \\
   DIRBE & F4.9 & 4.9   & 153 & 0.7\degr \\
   JP11 & M & 5.0   & 158 & -- \\
\hline
   GLIMPSE  & IRAC8 & 7.87   & 64.1 &  2\arcsec\\
   MSX  & A & 8.28  & 58.5 & 18.3\arcsec \\      
   AKARI  & S9W & 9.0   & 56.3 & 5.5\arcsec\\
   10$\mu$m-CAT. & 10$\mu$m & 10.0  & 30.0 &  -- \\
   JP11 & N & 10.2   & 43.0 & -- \\
  IRAS & F12 & 11.43   & 28.3 & 0.5\arcmin \\
   WISE & W3 & 11.56   & 31.7 & 6.5\arcsec \\
  DIRBE & F12 & 12.0   & 29.0 & 0.7\degr \\
   MSX  & C & 12.13  & 26.1 & 18.3\arcsec \\      
\hline                                   
\end{tabular}
\end{table}

In addition to the photometric measurements reported in the catalogues listed here above, 
we use flux estimations that we average for each MIR spectral band, reported in:
\begin{itemize}
    \item the VLTI/MIDI list of calibrator candidates (403 sources) for the SAAO/L and IRAS/12 filters ;
    \item the tables of Parameters and IR excesses of \textit{Gaia} DR1 stars \citep[Cat. J/MNRAS/471/770;][1.47 million sources covering the entire sky]{mcdonald17} for the WISE/W1, WISE/W2, WISE/W3, AKARI/S9W, and IRAS/12 filters; 
    \item the  blackbody model  for $\lambda$~=~3.5, 4.8, and 10.5~$\mu$m,  using the values of effective temperature and angular diameter reported in the VLTI/MIDI list and the \textit{Gaia} DR2;
    \item the complete set of Cohen's standards (435 sources)  for $\lambda$~=~10.7~$\mu$m, downloadable from the Gemini Observatory Website (with the link given in Appendix~\ref{Catalogs}).
\end{itemize}

    The values used to compute the final flux values with their statistical dispersions are reported in: 6 databases for the $L$-band; 8 databases for the $M$-band; and 12 databases  for the $N$-band.
 Table~\ref{table:nbval} gives the number of entries as a function of the number of flux values reported in each MIR band. In our catalogue, 93\% of the entries have at least 2 flux values reported in each band, while 4\% have at least 3 flux values in $L$ and $M$, 35\% in $N$. The maximum number of flux values effectively used are: 5 for the $L$-band, 6 for the $M$-band and 9 for the $N$-band. 
 
%
\begin{table}
\caption{Number of entries for each number of reported individual flux values for each band.}             
\label{table:nbval}      
\centering                          
\begin{tabular}{c | r r r }        
\hline\hline                 
\# of flux values & in L & in M & in N \\    
\hline                        
  0  & 102 & 100 & 107 \\      
   1  & 33\,683 & 33\,627  & 31\,781 \\
   2 & 412\,467 & 411\,345  & 270\,212 \\
  3  & 19\,058 & 19\,417 & 85\,130 \\     
   4  & 487 & 1\,028  & 65\,802 \\
   5 & 60 & 288  & 9\,179 \\
   6  & 0 & 52  & 3\,047 \\
   7 & 0 & 0  & 529 \\
8  & 0 & 0 & 64 \\     
 9  & 0 & 0 & 6 \\     
\hline                                   
\end{tabular}
\end{table}

It should be noted that the merging of individual flux data obtained with such a different angular resolution (from 1.7\arcsec~for GLIMPSE to 0.7\degr~for DIRBE) may cause source confusion. Flux values may be slightly overestimated with observing beams including numerous unresolved sources and their circumstellar environment. 

For the star 72~Leo taken as an example in Fig.~\ref{fig:viewer}, the median flux values (drawn as black squares) are $F_{L}$~=~327~Jy,  $F_{M}$~=~174~Jy, and  $F_{N}$~=~44~Jy, computed with 5 individual values for $L$, 6 values for $M$, and 8 values for $N$  (see Table~\ref{table:fluxes} for details). The $MAD$ values are 19~Jy for $L$, 37~Jy for $M$, and 11.5~Jy for $N$. We observe that the median values match the mean SED (spectral energy distribution), falling between the SEDs of the blackbody model with $\phi$~=~5.9~mas, $T_\mathrm{eff}$~=~3900~K (\textit{Gaia} DR2 values), and with $\phi$~=~5.8~mas, $T_\mathrm{eff}$~=~3600~K (MIDI values). The calculation of the mean and standard deviation of the flux would give: $\bar{F}_{L}$~=~281$\pm$105~Jy; $\bar{F}_{M}$~=~163$\pm$49~Jy; and $\bar{F}_{N}$~=~47$\pm$13~Jy.  Based on the $MAD$ assessment, the average uncertainty that we report in our catalogue is lower than the standard deviation $\sigma$, since outliers can heavily influence $\sigma$, while deviations of a small number of outliers are irrelevant in the $MAD$. Note that for a normal distribution, $MAD$ is related to $\sigma$ as $MAD~\sim~0.67\sigma$.

    Appendix~\ref{Catalogs} lists the complete set of databases used 
    to build our catalogue, with the number of entries reported in each of them and the references.

%
\begin{table}
\caption{Flux values reported in our catalogue for 72 Leo (-- indicates that no value is reported, ``BB'' stands for ``blackbody'') in the 3 bands: $L$ (top), $M$ (middle), and $N$ (bottom).}             
\label{table:fluxes}      
\centering                          
\begin{tabular}{l c c }        
\hline\hline                 
 Catalogue & Wavelength & Flux  \\    
                &  ($ \mu$m) & (Jy) \\
\hline                        
   WISE\_W1  & 3.35 & 95.1 \\      
   BB\_MIDI  & 3.5  & 307.2  \\
   BB\_\textit{Gaia}  & 3.5  & 328.1  \\
   JP11\_L & 3.5   & 349.5 \\
  DIRBE\_F3.5 & 3.5   & 326.6  \\
  GLIMPSE\_IRAC3.6 & 3.6   & -- \\
\hline
   MSX\_B1 & 4.29  & 78.9 \\      
   MSX\_B2 & 4.35    & 218.2 \\
   GLIMPSE\_IRAC4.5 & 4.5  & --\\
   WISE\_W2 & 4.6  & 173.1 \\
  BB\_MIDI  & 4.8  & 173.9  \\
   BB\_\textit{Gaia}  & 4.8  & 197.8  \\
   DIRBE\_F4.9 & 4.9   & 136.9 \\
   JP11\_M & 5.0   &  -- \\
\hline
   GLIMPSE\_IRAC8 & 7.87   & --\\
   MSX\_A & 8.28  & 60.9 \\      
   AKARI\_S9W & 9.0   & 68.5\\
   10$\mu$m-CAT & 10.0  & 49.5\\
   JP11\_N & 10.2   & -- \\
   BB\_MIDI  & 10.5  & 41.1  \\
   BB\_\textit{Gaia}  & 10.5  & 44.4  \\
  COHEN & 10.7 & -- \\
  IRAS\_F12 & 11.43   & -- \\
   WISE\_W3 & 11.56   & 33.0 \\
  DIRBE\_F12 & 12.0   & 34.2 \\
   MSX\_C & 12.13  & 31.8 \\      
\hline                                   
\end{tabular}
\end{table}

\section{ Identifying the stars with IR extended features}\label{section:IRflag}

    The JSDC reports a 3-bit flag , called "CalFlag", which identifies the stars that should not be
    used as interferometric calibrators because of: 
\begin{enumerate}
          \item the uncertain estimation of their
    reconstructed angular diameter, with $\chi^2$~>~5 \citep[see Appendix A.2 of][]{chelli16};
         \item their close binarity, with $\varepsilon$~<~1\arcsec reported in WDS;
          \item their suspect Object Type in SIMBAD which signals a possible
              binarity or pulsating stars.
\end{enumerate}
   Beside "CalFlag", we define a new 3-bit flag, called "IRflag". This new flag identifies the stars that show probable extended features at MIR wavelengths revealed by their photometric excess,  extent, and/or variability.

\subsection{Tagging the IR-excess}\label{section:IRexc}
The first bit of IRflag is set if the star shows an IR-excess, identified 
              thanks to the values of: 
\begin{itemize} 
             \item the [K-W4] color index\footnote{[W4] means the magnitude at 22.1~$\mu$m reported in the WISE All-Sky Data Release (W4 filter)}  defined in 3 different
                [J-H] parts \citep[according to][]{wu13}: (i) [K-W4]~$\ge$~0.26 for [J-H]~$\le$~0.1; (ii) [K-W4]~$\ge$~0.21 for 0.1~<~[J-H]~$\le$~0.3; and
           (iii) [K-W4]~$\ge$~0.22 for [J-H]~>~0.3 ;
          \item the overall MIR-excess statistic X$_{\rm MIR}$ reported by \citet{mcdonald17} for a large sample of Tycho-2 and
\textit{Hipparcos} stars with distances from \textit{Gaia} DR1, following the IR-excess criterion: X$_{\rm MIR}$~>~1.15+A$_{V}$/100, where A$_{V}$ is the optical extinction.
\end{itemize} 
It should be noted, however, that sources tagged as IR-excess stars in our catalogue might also include good candidates wrongly tagged with these criteria, since the IR-excess could also be from nearby bright stars or background sources.
 
\subsection{Tagging the IR-extent}\label{section:IRext}
The second bit of IRflag is set if the star is far-extended in the IR (at angular scale of several arcseconds),
              indicated by the extent flags reported in the 2 catalogues WISE/AllWISE and AKARI. We consider the source as possibly non-extended if it has: 
\begin{itemize} 
              \item the WISE extent flag~=~0, meaning that: (i) the source shape is consistent with a point source (FWHM~=~6.1\arcsec in $L$ and $M$, 6.5\arcsec in $N$), with a goodness-of-fit value of the photometric profile fit~<~3 in all bands; and (ii) that the source is not associated with or superimposed on a NIR source of the 2MASS Extended Source Catalog; or
              \item the AKARI extent flag~=~0. Unity flag means that the average of the radius along the major and minor axes of the source extent estimated with SExtractor is~>~15.6\arcsec, more extended than the PSF in the $S9W$ band (FWHM~=~5.5\arcsec).
\end{itemize}
It should be noted, however, that very bright point sources  of our catalogue might also appear as extended with these criteria. Moreover, the lack of information on circumstellar extent at the subarcsecond scale (below 1\arcsec) does not warranty the source to be free from a close circumstellar environment (better revealed by the IR-excess).

\subsection{Tagging the MIR variability}\label{section:IRvar}

The third (and last) bit of IRflag is set if the star is a likely variable in the MIR, identified
              by the variability flags reported in the WISE/AllWISE catalogues, the MSX6C Infrared Point Source
Catalogue, the IRAS Point
      Sources Catalogue, and the 10-micron Catalog. We consider the source as a likely variable if it:
\begin{itemize} 
   \item is tagged as a likely variable in at least one of the WISE filters W1, W2, or W3 (variability flag of "7" or "8"); or
   \item has the variability flag~=~1 in at least one of the MSX filters B1, B2, C, or A; or
   \item has the likelihood of variability~>~90\% in the IRAS/12 filter; or
   \item is tagged as a variable star for $\lambda$~=~10~$\mu$m.
\end{itemize}
Stars fulfilling none of those criteria listed hereabove are unlikely variable. It should be noted, however, that sources with false-positive variability reported in our catalogue might also be considered as likely variable with those criteria.

%
\begin{table}
\caption{Number of entries for each binaries subset  (see Sect.~\ref{bin} for the definition of the subsets, based on the WDS).}             
\label{table:binaries}      
\centering                          
\begin{tabular}{l | r   }        
\hline\hline                 
 Subset & \# of entries \\
\hline                        
Wide binaries  &  19\,895 \\
Intermediate binaries  & 4\,478  \\
Close binaries  &  3\,630 \\
\hline
\end{tabular}
\end{table}

\section{Results} \label{section:results}

Appendix~\ref{append:TabCol} gives the meaning of the columns of our catalogue. The description is presented as a three-column table with the following elements: 
\begin{itemize} 
   \item a label or column header;
   \item the unit in which the value is expressed; and 
   \item a short explanation of the contents of the column.
\end{itemize}
Our catalogue is downlable at https://matisse.oca.eu/foswiki\footnote{https://matisse.oca.eu/foswiki/pub/Main/TheMid-infraredStellarDiametersAndFluxesCompilationCatalogue(MDFC)/mdfc-v10.zip}. It contains a total of 465\,857 entries covering the entire sky, including 201\,200 "pure" calibrators (with null CalFlag and IRflag), and 28\,003 binaries reported in the WDS. Only 102 entries have no individual flux measurement or even estimate reported in the $L$-band, 100 in the $M$-band, and 107 in the $N$-band, while 88 entries have no flux value reported in any of these 3 bands. The number of sources visible from the ESO-Paranal Observatory reaches 371\,333, i.e. 80\% of the catalogue entries, including 156\,602 "pure" calibrators.

	\subsection{Sky coverage}

   \begin{figure*}
   \centering
  \includegraphics[trim = 0cm 0cm 0cm 0cm, clip, width=14cm]{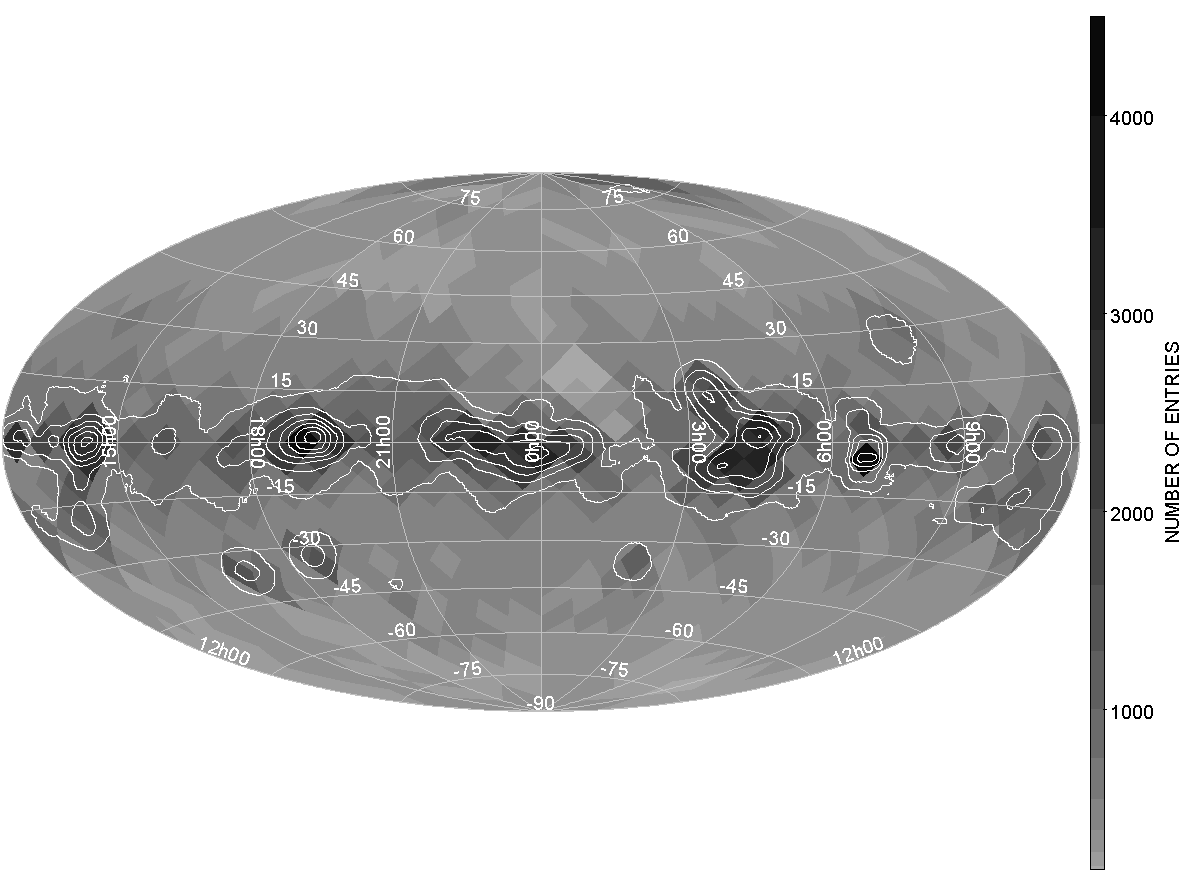}
      \caption{All-sky density map of the whole catalogue, in galactic coordinates.}
         \label{fig:SkyCov}
   \end{figure*}

Figure~\ref{fig:SkyCov} shows the all-sky density map of all the entries of the catalogue in galactic coordinates.The overdensities are distributed along the Milky Way, highlighted by the contour lines. The density in galactic longitude is higher than 14\,700 entries per bin of 1 hour.

	\subsection{Spectral and luminosity class distribution}

   \begin{figure}
   \centering
  \includegraphics[trim = 0cm 0cm 0cm 0cm, clip, width=\hsize]{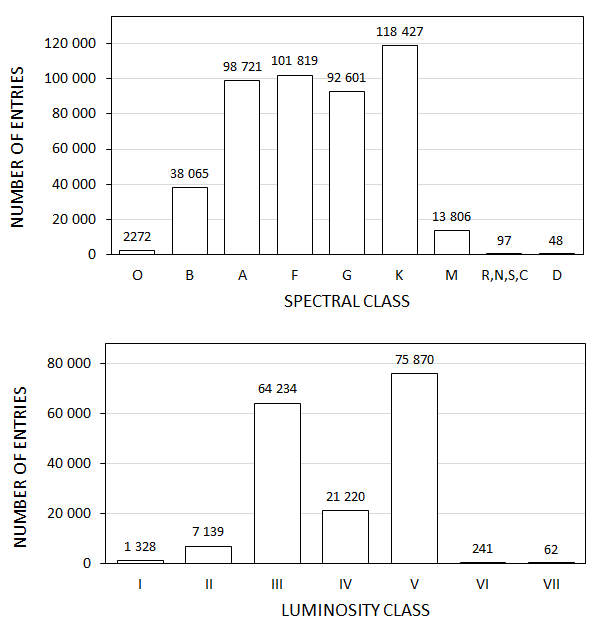}
      \caption{Distribution in spectral (top) and luminosity (bottom) class for the entries of our catalogue.
              }
         \label{fig:SpType}
   \end{figure}

Figure~\ref{fig:SpType} shows the distribution of spectral and luminosity classes of the entries of the catalogue. 
About 9\% of catalogue entries have spectral class O or B; 88\% have spectral class A, F, G, or K; 3\% have spectral class M, R, N, S, C, or D. Only one entry of the catalogue (BD+47~2769, an eruptive variable star) has no spectral type reported in SIMBAD. 
About 2\% of catalogue entries have luminosity class I or II; 35\% have luminosity class III, IV, or V; 0.07\% have luminosity class VI or VII; and 63\% of the entries have no luminosity class reported in SIMBAD.

	\subsection{Distance distribution}

   \begin{figure}
   \centering
  \includegraphics[trim = 0cm 0cm 0cm 0cm, clip, width=\hsize]{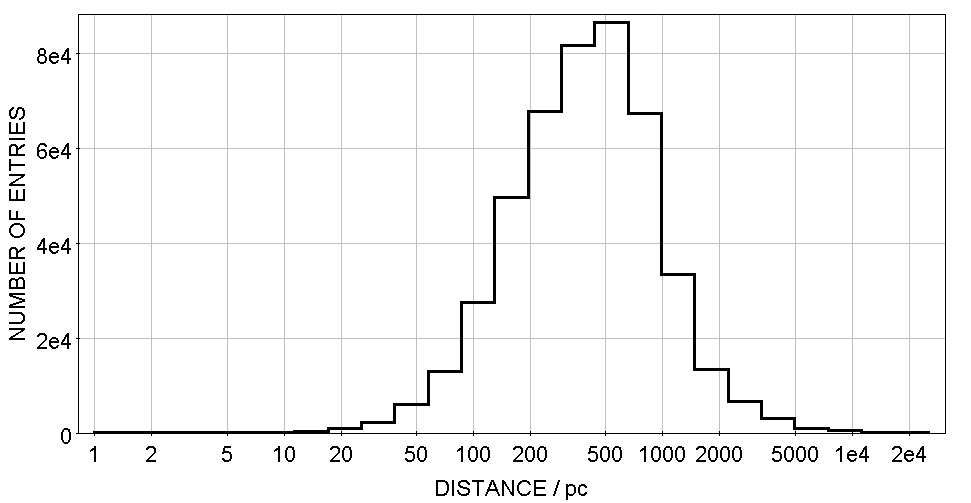}
      \caption{Distribution in geometric distance computed by \citet[][]{bailerjones18} for the entries of our catalogue.
              }
         \label{fig:DistDistrib}
   \end{figure}

Figure~\ref{fig:DistDistrib} shows the distribution of the geometric distance from \citet{bailerjones18} :
\begin{itemize}
\item lower quartile distance:~215~pc;
\item median distance:~400~pc;
\item upper quartile distance:~690~pc. 
\end{itemize} 
About 6\% of the entries have $D$~<~100~pc, and only 1\% have $D$~<~43~pc. The most distant source of the catalogue is located at 18.4~kpc (BM~VII~9; A1IIIe Spectral Type). We want to stress that the distance values at the kpc scale from the \textit{Gaia} DR2 must be used with caution \citep[see e.g.][]{luri18}.

	\subsection{Binaries distribution}\label{bin}

   \begin{figure}
   \centering
  \includegraphics[trim = 0cm 0cm 0cm 0cm, clip, width=\hsize]{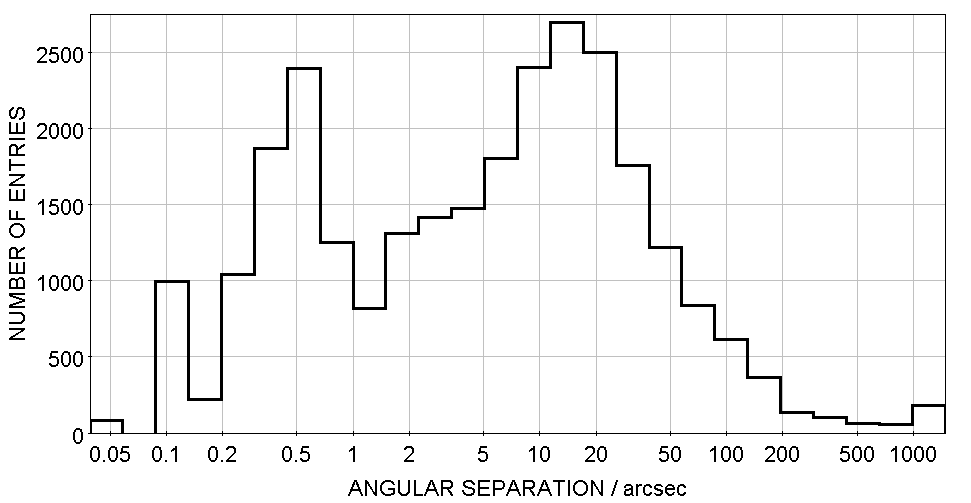}
      \caption{Distribution in angular separation of the sources of our catalog identified as binaries in the WDS.
              }
         \label{fig:SepDistrib}
   \end{figure}

%
\begin{table}
\caption{Number of entries for each angular diameter subset  (see Sect.~\ref{diam} for the definition of the subsets, based on the JSDC).}             
\label{table:resolved}      
\centering                          
\begin{tabular}{l | r   }        
\hline\hline                 
 Subset  & \# of entries \\
\hline                        
 Large  &  5 \\
 Medium-size  & 74  \\
 Small  & 762  \\
 Point-like  & 464\,763  \\
\hline
\end{tabular}
\end{table}

Figure~\ref{fig:SepDistrib} shows the distribution in angular separation ($\varepsilon$) of the 28\,003 astrometric binaries reported in the catalogue from the WDS: 
\begin{itemize}
\item lower quartile separation:~2\arcsec;
\item median separation:~21\arcsec;
\item upper quartile separation:~77\arcsec. 
\end{itemize} 
Only 2\% of the entries have $\varepsilon$~<~0.1\arcsec.

According to their angular separation, we divide the astrometric binaries contained in our catalogue in 3 different subsets:
\begin{enumerate}
 \item The wide binaries: $\varepsilon$~$\ge$~1\arcsec.
 \item The intermediate binaries: 0.4\arcsec~$\le$~$\varepsilon$ < 1\arcsec.
 \item The close binaries: $\varepsilon$~<~0.4\arcsec.
\end{enumerate}
Table~\ref{table:binaries} gives the number of entries for each subclass of the binaries.

	\subsection{Angular diameter distribution}\label{diam}

   \begin{figure}
   \centering
  \includegraphics[trim = 0cm 0cm 0cm 0cm, clip, width=\hsize]{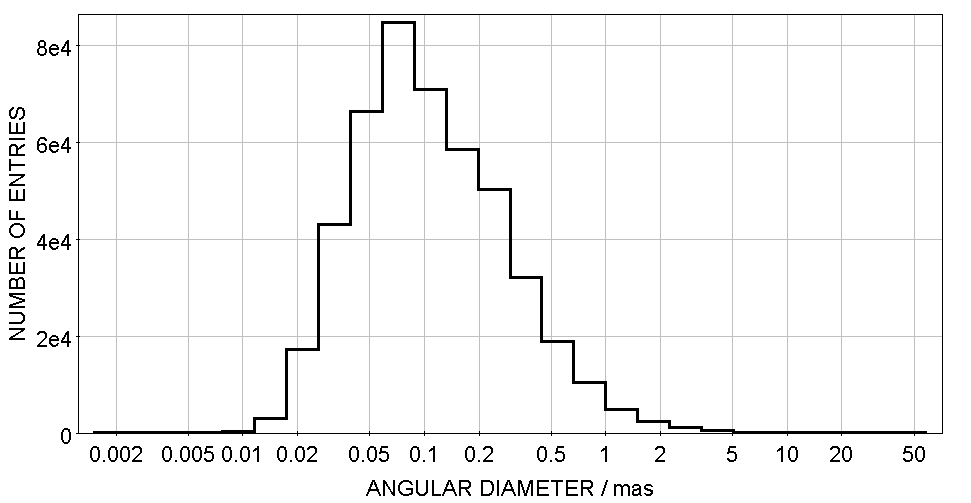}
      \caption{Distribution in angular diameter reported in the JSDC for the entries of our catalogue.
              }
         \label{fig:LddDistrib}
   \end{figure}

   \begin{figure*}
   \centering
  \includegraphics[trim = 0cm 0cm 0cm 0cm, clip, width=16cm]{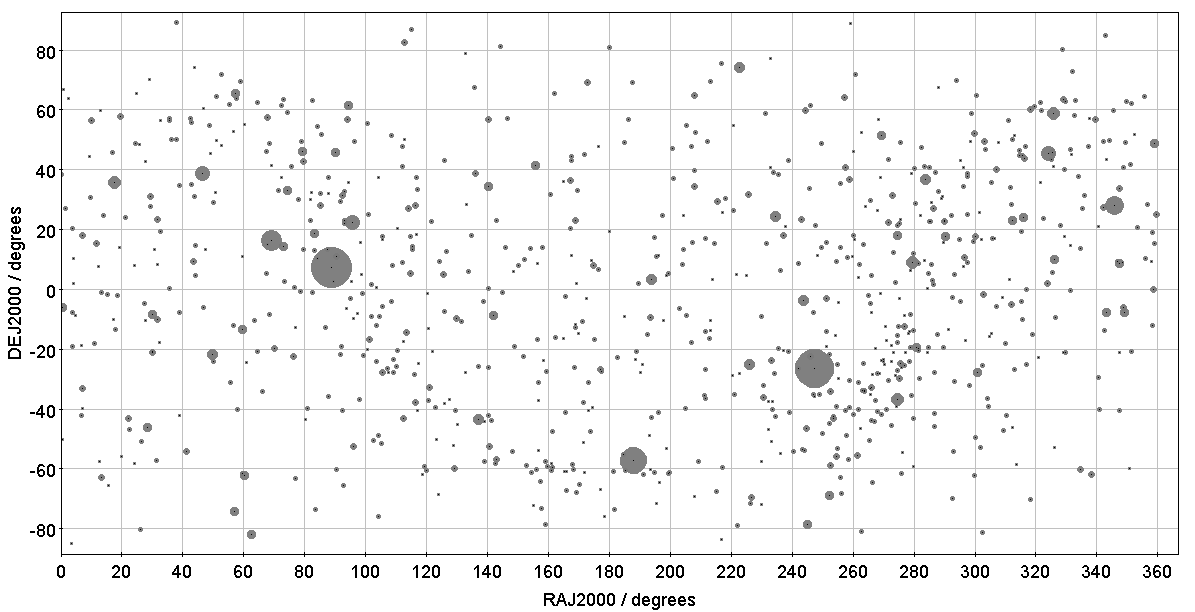}
      \caption{All-sky coverage of the 841 "resolved" sources contained in our catalogue ($\phi$~>~3~mas). The size of the dots is proportional to the value of the LDD diameter reported in the JSDC. }
         \label{fig:DiamCov}
   \end{figure*}

Figure~\ref{fig:LddDistrib} shows the distribution of the LDD diameter ($\phi$) computed by \citet{chelli16} and reported in the JSDC:
\begin{itemize}
\item lower quartile diameter:~0.05~mas;
\item median diameter:~0.1~mas;
\item upper quartile diameter:~0.2~mas. 
\end{itemize}
Only 2\% of the entries have $\phi$~>~1~mas.

According to their angular diameter, we divide the stellar sources contained in our catalogue into 4 different subsets:
\begin{enumerate}
 \item the "large" sources, fully resolved in $L$ and $N$ with a 130-m projected interferometric baselength: $\phi$~$\ge$~20 mas; 
 \item  the "medium-size" sources, fully resolved in $L$ and partially resolved in $N$: 7~$\le$~$\phi$~<~20 mas;
 \item  the "small" sources, partially resolved in $L$ and unresolved in $N$: 3~$\le$~$\phi$~<~7 mas;
 \item the "point-like" sources, unresolved in $L$ and $N$: $\phi$~<~3 mas.
\end{enumerate}

Table~\ref{table:resolved} gives the number of entries for each subclass of angular diameter. Only 841 entries are "resolved" sources ($\phi$~>~3~mas), and 252 have no angular diameter estimate reported in the JSDC. The 5 "large" sources are: $\alpha$~Ori ($\phi$~$\sim$~45~mas); $\alpha$~Sco ($\phi$~$\sim$~42~mas); $\gamma$~Cru ($\phi$~$\sim$~28~mas); $\alpha$~Tau ($\phi$~$\sim$~23~mas); and $\beta$~Peg ($\phi$~$\sim$~20~mas). Figure~\ref{fig:DiamCov} shows the all-sky coverage of these 841 sources. The size of the dots is proportional to $\phi$.

	\subsection{MIR flux distribution}\label{flux}
   \begin{figure}
   \centering
  \includegraphics[trim = 0cm 0cm 0cm 0cm, clip, width=\hsize]{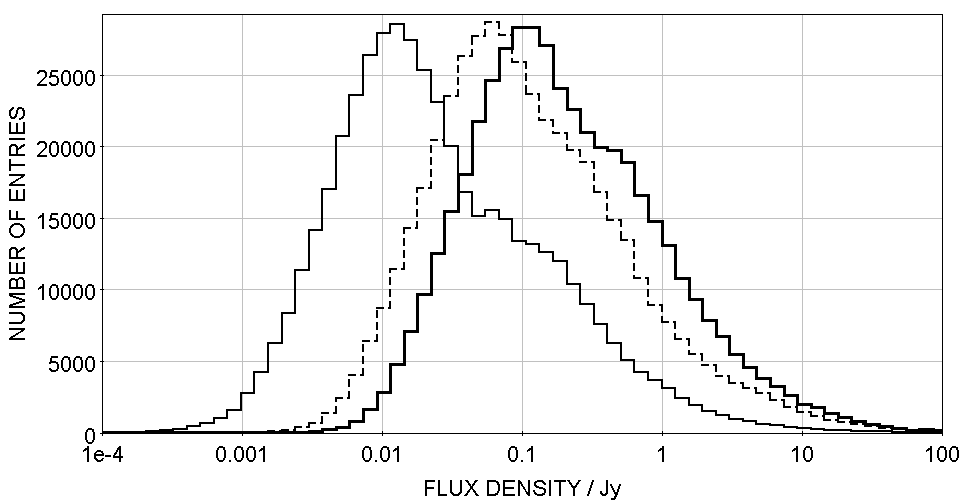}
      \caption{Distributions in flux density for the bands: $L$ (thick solid line), $M$ (thin dotted line), and $N$ (thin solid line). 
              }
         \label{fig:FlxDistrib}
   \end{figure}

   \begin{figure}
   \centering
  \includegraphics[trim = 0cm 0cm 0cm 0cm, clip, width=\hsize]{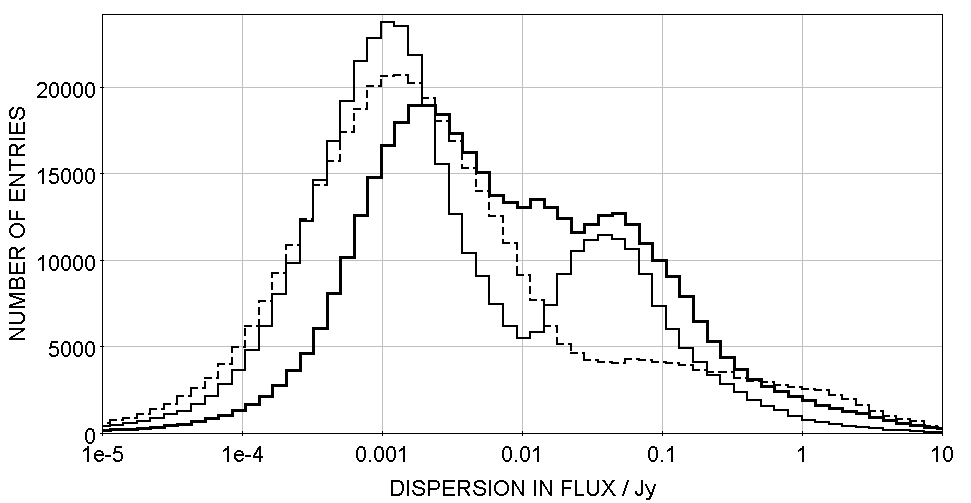}
      \caption{Distributions in flux density dispersion for the bands: $L$ (thick solid line), $M$ (thin dotted line), and $N$ (thin solid line). 
              }
         \label{fig:MadDistrib}
   \end{figure}

   \begin{figure*}
   \centering
  \includegraphics[trim = 0cm 0cm 0cm 0cm, clip, width=16cm]{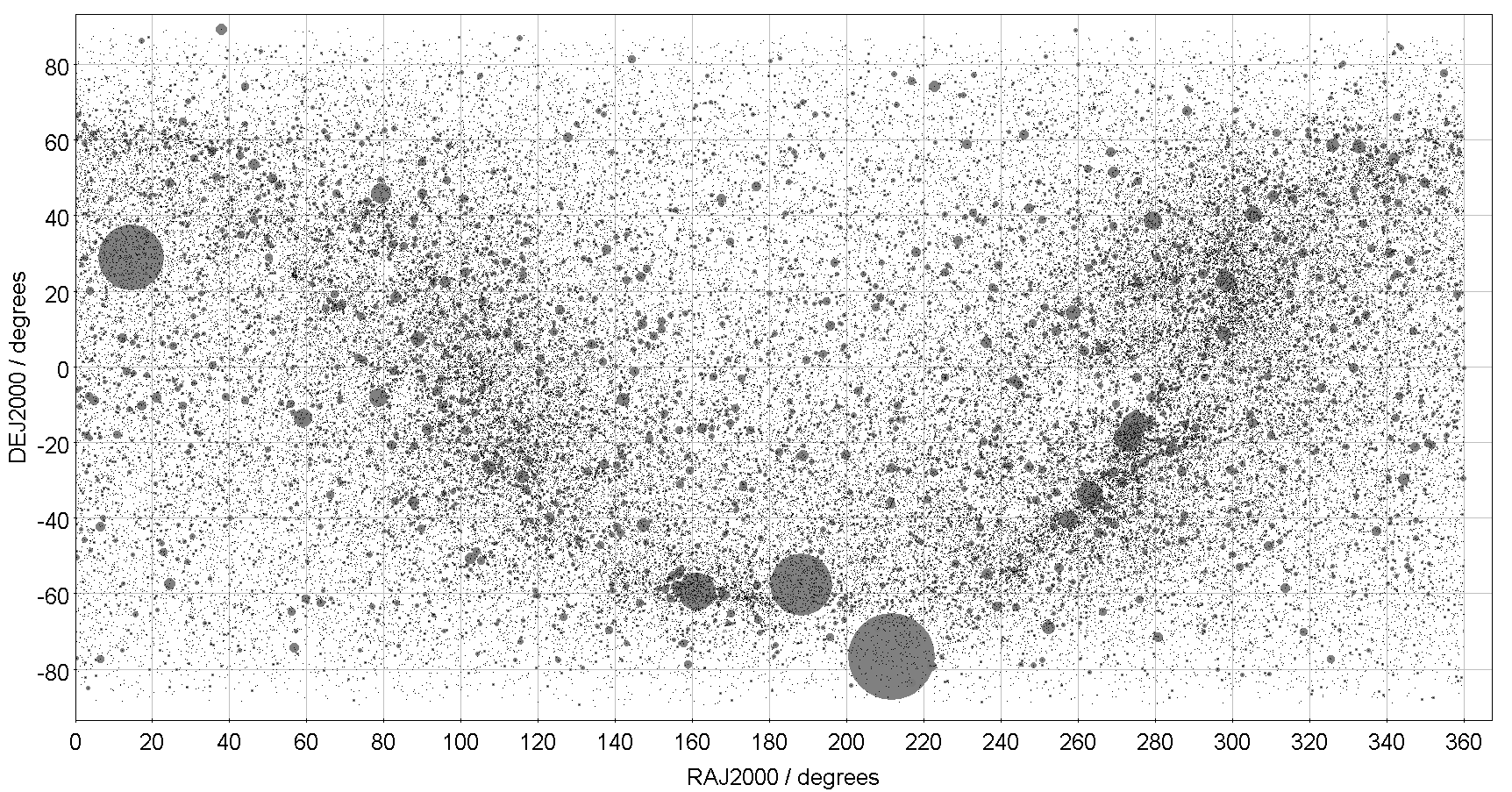}
      \caption{All-sky coverage of the 101\,431 "UT-bright" sources for the $L$-band ($C_{ L}$~>~0.7~Jy). The size of the dots is proportional to $C_{ L}$.}
         \label{fig:BrightCovL}
   \end{figure*}

   \begin{figure*}
   \centering
  \includegraphics[trim = 0cm 0cm 0cm 0cm, clip, width=16cm]{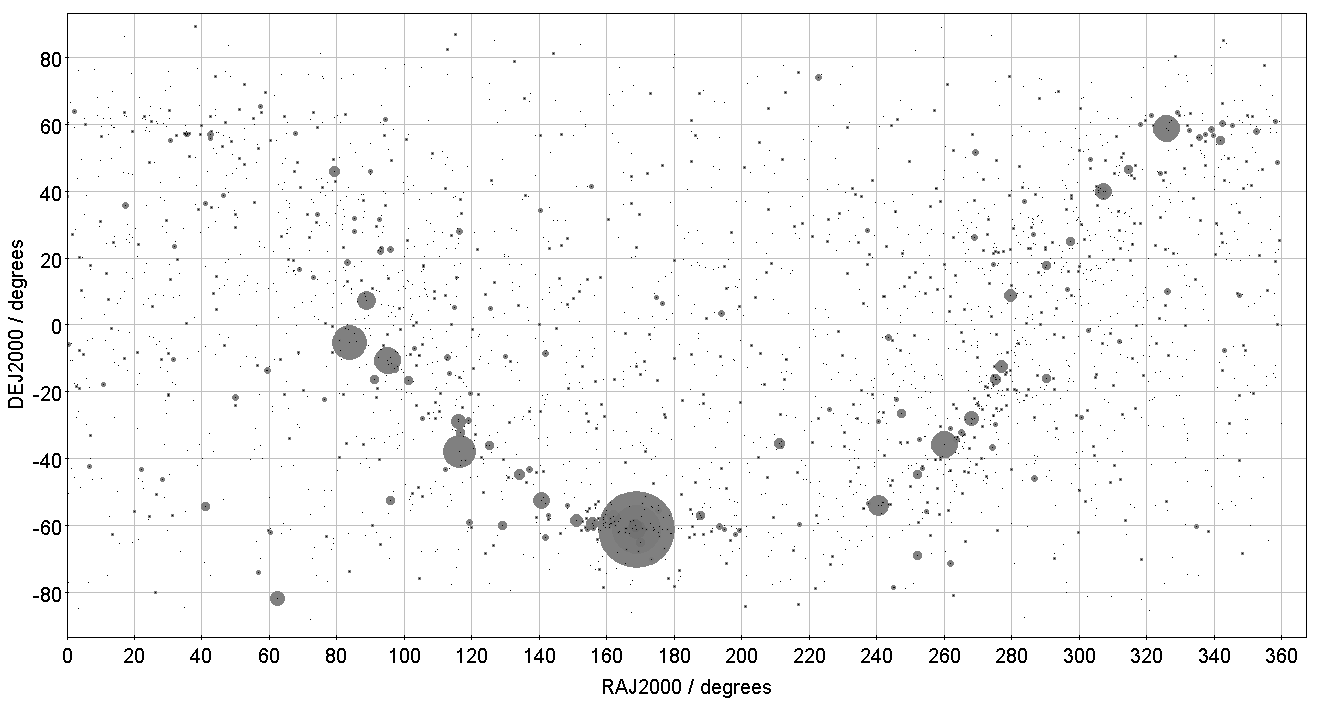}
      \caption{All-sky coverage of the 2\,605 "UT-bright" sources for the $N$-band ($F_{ N}$~>~5~Jy and  $C_{ N}$~>~3~Jy). The size of the dots is proportional to $C_{ N}$.}
         \label{fig:BrightCovN}
   \end{figure*}

%
\begin{table*}
\caption{Brightness subsets derived from the preliminary sensitivity performance of VLTI/MATISSE obtained during the commissioning phase.  The values of flux $F$ and correlated flux $C$ in the $L$- and $N$-bands are in Jy.}             
\label{table:categories}      
\centering                          
\begin{tabular}{l | c c | c c | c c | c c }        
\hline\hline                 
 & \multicolumn{4}{c |} {ATs}  & \multicolumn{4}{c} {UTs} \\
 Subset  &   \multicolumn{1}{c }{$F_L$}  & \multicolumn{1}{c |}{$C_L$} & \multicolumn{1}{c}{$F_N$} &  \multicolumn{1}{c |}{$C_N$}  &   \multicolumn{1}{c }{$F_L$}  & \multicolumn{1}{c |}{$C_L$} & \multicolumn{1}{c}{$F_N$} &  \multicolumn{1}{c}{$C_N$} \\
\hline                        
Bright & - & >~10  & - &  >~90  & - & >~0.7 & >~5 &  >~3\\
Medium-bright & 1.5-10 & 1-10  & 28-90 & 20-90 & 0.1-0.7 &  0.07-0.7 & 1.5-5 & 0.7-3\\
Faint & - & 0.2-1 & - & 4-20 & - & 0.015-0.07 & - & 0.15-0.7\\
Undetectable & - & <~0.2 & - & <~4 & - & <~0.015 & - & <~0.15\\
\hline\end{tabular}
\end{table*}

%
\begin{table}
\caption{Number of entries for each brightness class in the $L$- and $N$-bands, using the ATs and the UTs.}             
\label{table:corflux}      
\centering                          
\begin{tabular}{l | r r | r r }        
\hline\hline                 
 Subset & \multicolumn{2}{c |} {ATs}  & \multicolumn{2}{c} {UTs} \\
 &   \multicolumn{1}{c}{$L$}  & \multicolumn{1}{c |}{$N$} &  \multicolumn{1}{c}{$L$}  &\multicolumn{1}{c}{$N$}\\
\hline                        
Bright & 8\,221 & 44 & 101\,431 & 2\,605\\
Medium-bright & 46\,298 & 253 & 197\,707 & 5\,011\\
Faint & 138\,793 & 2\,857 & 110\,322 & 58\,457\\
Undetectable & 249\,976 & 462\,220 & 12\,120 & 386\,358\\
\hline\end{tabular}
\end{table}

Figure~\ref{fig:FlxDistrib} shows the distributions in median flux for the $L$-, $M$-, and $N$-bands:
 \begin{itemize}
\item lower quartile flux:~0.07~Jy ($L $), 0.04~Jy ($M$), and 0.01~Jy ($N$);
\item median flux:~0.17~Jy ($L$), 0.09~Jy ($M$), and 0.02~Jy ($N$);
\item upper quartile flux:~0.6~Jy ($L$), 0.3~Jy ($M$), and 0.1~Jy ($N$). 
\end{itemize} 
About 36\% of the entries have $F_{L}$~<~0.1~Jy (51\% for $M$, 78\% for $N$).  Figure~\ref{fig:MadDistrib} shows the distributions in flux dispersion for the $L$-, $M$-, and $N$-bands. The mean ratio dispersion to flux value (relative dispersion, $rdisp$) is 6\% for the $L$- and $M$-bands, and 15\% for the $N$-band:
 \begin{itemize}
\item lower quartile relative dispersion:~2\% ($L $), 1\% ($M$), and 6\% ($N$);
\item median relative dispersion:~4\% ($L$), 2\% ($M$), and 10\% ($N$);
\item upper quartile relative dispersion:~8\% ($L$), 6\% ($M$), and 23\% ($N$). 
\end{itemize} 

We divide the sources contained in our catalogue according to their median flux $F$ and correlated flux $C$ in the $L$- and $N$-bands, assuming the UD model and the 130-m projected baselength. The relation between $C$ and $F$ is
\begin{equation}
C/F~=~V~=~2~\frac{\lvert{J_{1}(\pi \phi B / \lambda_{c})}\rvert}{\pi \phi B / \lambda_{c}}, 
\label{UDvis} 
\end{equation}
where $B$~=~130~m, $\lambda_{c}$~=~3.5, 4.8 and 10.5~$\mu$m for the $L$-, $M$-, and $N$-bands  respectively, and $\phi$ is the UD diameter for the considered band (reported in the JSDC). We define the 4 following subsets derived from the preliminary sensitivity performance of VLTI/MATISSE obtained during the commissioning phase, with corresponding ranges in flux and correlated flux given in Table~\ref{table:categories}: 
\begin{enumerate}
\item "Bright" sources, for which the calibration quality in visibility is independent of the flux value.
\item "Medium-bright" sources, for which the calibration quality in visibility depends on the flux value.
\item "Faint" sources, which remain observable and useable for coherent flux observation.
\item "Undetectable" sources, for which no fringe detection is achieved with current standard observations.
\end{enumerate}

Table~\ref{table:corflux} gives the number of entries for each brightness subset, using the ATs and the UTs in the $L$- and $N$-bands . Figures~\ref{fig:BrightCovL} and \ref{fig:BrightCovN} show the all-sky coverage of the "UT-bright" sources for the $L$- and $N$-bands  respectively. In both figures, the size of the dots is proportional to the correlated flux in the considered band.

	\subsection{Flags distribution and "pure" calibrators}\label{flags}

%
\begin{table}
\caption{Truth table of CalFlag and number of entries. OType is the Object Type in SIMBAD.}             
\label{table:CalFlag}      
\centering                          
\begin{tabular}{c c c c | r }        
\hline\hline                 
CalFlag & $\chi^2 > 5$ & $\varepsilon < 1\arcsec$ & Bad  & \# of entries \\    
 &  &  &  OType &  \\    
\hline                        
  "0"  & no & no & no &  450\,921\\      
   "1"  & yes & no   & no & 371 \\
   "2" & no & yes  & no & 8\,090 \\
  "3"  & yes & yes & no & 31 \\     
   "4"  & no & no   & yes & 5\,965\\
   "5" & yes & no  & yes & 66 \\
   "6"  & no & yes   & yes &  404\\
   "7" & yes & yes  & yes &  9\\
\hline                                   
\end{tabular}
\end{table}

%
\begin{table}
\caption{Truth table of IRflag and number of entries for each value of the flag. }             
\label{table:IRflag}      
\centering                          
\begin{tabular}{c c c c | r }        
\hline\hline                 
IRflag & IR-excess & IR-extent &  MIR-var. & \# of entries \\    
\hline                        
  "0"  & no & no & no &  207\,434\\      
   "1"  & yes & no   & no & 162\,899 \\
   "2" & no & yes  & no & 30\,570 \\
  "3"  & yes & yes & no & 36\,020 \\     
   "4"  & no & no   & yes$^{a}$ &  4\,815\\
   "5" & yes & no  & yes$^{a}$  & 4\,345 \\
   "6"  & no & yes   & yes$^{a}$ &  7\,075\\
   "7" & yes & yes  & yes$^{a}$ &  12\,699\\
\hline                                   
\end{tabular}
\\
\begin{flushleft}
{
$^{a}${The MIR variability is defined in Sect.~\ref{section:IRvar}. The likely variable sources must fulfill at least one of the following criteria: (i) variability flag~>~6  in at least one of the WISE filters W1, W2, or W3; (ii) variability flag~=~1  in at least one of the MSX filters B1, B2, C, or A; (iii) likelihood of variability~>~90\% in the IRAS/12 filter; (iv) tagged as variable in the 10-micron Catalog.}
}
\end{flushleft}
\end{table}

%
\begin{table}
\caption{Crossed distribution of CalFlag and IRflag.}             
\label{table:CalIR}      
\centering                          
\begin{tabular}{ c | r r r r r r r r }        
\hline \hline
IR- &  \multicolumn{8}{c}{CalFlag}\\
flag &  \multicolumn{1}{c}{"0"} & \multicolumn{1}{c}{"1"} & \multicolumn{1}{c}{"2"} & \multicolumn{1}{c}{"3"} & \multicolumn{1}{c}{"4"} & \multicolumn{1}{c}{"5"} & \multicolumn{1}{c}{"6"} & \multicolumn{1}{c}{"7"}    \\
\hline                        

 "0" & 201\,200 & 16  &  4\,034 & 0 & 2\,019 & 0 & 165 & 0  \\
"1" &  158\,992 & 146 & 2\,431 & 6 & 1\,269 & 11 & 44 & 0  \\
"2" &  29\,412 & 21 & 584 & 1 & 483  & 2 & 64 & 3  \\
"3" &  34\,875 &  75 & 541 & 9 & 462 & 20 & 35 & 3  \\
"4" & 4\,434 & 3 & 95 & 0 & 271 & 0 & 12 & 0  \\
"5" &  3\,752 & 23 &  76 &  1 & 482 & 6 & 5 & 0  \\
"6" & 6\,495 & 32 & 141 & 5 & 340 & 6 & 54 & 2  \\
"7" & 11\,761 &  55 & 188  & 9 & 639 & 21 & 25 & 1  \\
\hline
\end{tabular}
\end{table}

           Table~\ref{table:CalFlag} gives the truth table showing the possible values of CalFlag, with the number of entries corresponding to each value. Let us recall that CalFlag, reported in the JSDC (for detail see the description provided in the CDS/VizieR database for the II/346 JSDC catalogue), is a 3-bit flag taking values ranging from "0" to "7":
\begin{itemize}
          \item bit 1 is set if the chi-square associated with the
              reconstructed log diameter is >~5;
         \item bit 2 is set if the star is a known double in WDS 
              with separation <~1\arcsec;
          \item bit 3 is set if the star is, according to its SIMBAD's object
              type, a known spectroscopic binary, an Algol type star, a pulsating star, etc.
\end{itemize}
Although none of these flag values prevent the LDD diameter estimate to be accurate, they imply some caution in choosing this star as a calibrator star for OLBI.  

Table~\ref{table:IRflag} gives the truth table showing the possible values of  IRflag, with the number of entries corresponding to each value.  Let us recall that our new IRflag is also a 3-bit flag taking values ranging from "0" to "7":
\begin{itemize}
         \item bit 1 is set if the star shows an IR excess, identified thanks to the [K-W4] and [J-H] color indexes, and the overall MIR excess statistic X$_{\rm MIR}$ computed from \textit{Gaia} DR1;
         \item bit 2 is set if the star is extended in the IR, indicated by the extent flags reported in the WISE/AllWISE and AKARI catalogues;
          \item bit 3 is set if the star is a likely variable in the MIR, identified by the variability flags reported in the WISE/AllWISE catalogues, the MSX6C Infrared Point Source Catalogue, the IRAS PSC, and the 10-micron Catalog.
\end{itemize}

Table~\ref{table:CalIR} gives the number of entries of the catalogue for each value of the pair (CalFlag;IRflag).

We consider those entries of the catalogue as MIR interferometric calibrators which:
\begin{itemize} 
\item have a reliable angular diameter estimate;
\item are single stars or binaries with $\varepsilon$~>~1\arcsec;
\item have a favorable Object Type;
\item show no IR-excess;
\item show no IR-extent; and
\item is unlikely variable in the MIR.
\end{itemize} 

We find 201\,200 entries (43\% of the total number of entries of the catalogue) that satisfy these 6 conditions (i.e. with null CalFlag and IRflag). We consider them as
"pure" calibrators useable by high angular resolution instruments operating in the MIR. 
Strictly speaking, stars that do not fulfill this list of criteria should be kept with caution if used as interferometric calibrators. 
Finding stars fulfilling these criteria does not ensure them 
to be undoubted calibrators but potential ones only. For this reason, we 
strongly suggest to use more than one calibrator associated to each science target for the purpose of interferometric calibration. 

\section{Selecting calibrators for observations with MATISSE } \label{section:calib}

MATISSE is the second generation spectro-interferometer of the VLTI, designed to observe in the $L$-, $M$- and $N$-bands. Preparing the commissioning and the science observations with MATISSE triggered the work described in this paper.
MATISSE has several types of spectro-interferometric observables that can be used for model fitting and for image reconstruction. All these observables must be calibrated using a target for which their value is known, ideally a point source, or at least a disk with known diameter. This already excludes the binaries and the targets with IR excess that are unlikely to be well defined disks. The other parameters of a calibrator, such as the flux, the coherent flux and the angular proximity with the science target have to be selected with different criteria for the different observables and for the observation spectral band. The ideal calibrator has the same flux and the same coherent flux in $L$ and $N$ and is observed through the same atmospheric conditions, which means that its angular distance with the science target is small (less than a few degrees) and the "Cal-Sci" cycle is fast. In practice such a calibrator almost never exists. So, it might be necessary to use several calibrators to fulfill different constraints. In the following we try to propose criteria allowing the number of calibrators to be minimised and therefore the telescope time spent for the science target itself to be maximised.

The full calibration procedure of all MATISSE observables is a complex issue as there are several possible calibration strategies that can be adapted to the MATISSE observable favored by the user. A full discussion is beyond the scope of this paper and we give here only some global indications to ease the choice of the relevant calibrators.

The first observable of MATISSE is the coherent flux in each spectral channel $C_{ij}~=~V_{ij}~R_{ij}~\sqrt{ S_{i}~S_{j}}$ where $V_{ij}$ is the source visibility, $R_{ij}$ is the instrumental response in visibility, and  $S_{i}$ is the contribution of beam $i$ to the flux in the interferometric way. In the terminology of interferometric instruments $S_{i}$ is called the "photometry". The photometry can be written as $S_{i}~=~T_{i}~\eta_{i}~F$ where $T_{i}$ is the fixed and calibrated transmission of the instrument in the beam $i$, $\eta_{i}$ is a time-variable coupling efficiency that depends on the atmosphere and instrument variations through the Strehl ratio and the image jitter, and $F$ is the absolute flux of the source. 

An estimate of the response in visibility is provided by the observation of a reference star (calibrator), assuming no change of the instrumental response between the observation of the science target and its calibrator. Thus the coherent flux of the science target  in each spectral channel can be written as
\begin{equation}
  C_{\mathrm{sci},ij}~=~V_{\mathrm{sci},ij}~R_{ij}~\sqrt{ S_{\mathrm{sci},i}~S_{\mathrm{sci},j}}~=~\frac{V_{\mathrm{sci},ij}}{V_{\mathrm{cal},ij}}~C_{\mathrm{cal},ij}~\sqrt{\frac{S_{\mathrm{sci},i}~S_{\mathrm{sci},j}}{ S_{\mathrm{cal},i}~S_{\mathrm{cal},j}}}, \label{Cstar} 
\end{equation}
where the $_{\mathrm{sci}}$ subscript stands for the science target, and $_{\mathrm{cal}}$ for the calibrator.

	\subsection{Absolute visibility}

The most usual interferometric observable is the absolute visibility  in each spectral channel, defined according to Eq.~(\ref{Cstar}) as

\begin{equation}
  V_{\mathrm{sci},ij}~=~\frac{C_{\mathrm{sci},ij}}{C_{\mathrm{cal},ij}}~V_{\mathrm{cal},ij}~\sqrt{\frac{S_{\mathrm{cal},i}~S_{\mathrm{cal},j}}{ S_{\mathrm{sci},i}~S_{\mathrm{sci},j}}}. \label{Vstar} 
\end{equation}
Computing the ratio $(S_{\mathrm{cal},i}~S_{\mathrm{cal},j})~/~(S_{\mathrm{sci},i}~S_{\mathrm{sci},j})$ is the goal of the photometric calibration of MATISSE. The photometric measures $S_{i}$ are deduced by MATISSE itself using different parts of the detector or immediately after the interferometric measures. In both cases we introduce photometric errors that can bias the absolute visibility. To minimize this bias, one must use a photometric calibrator with the same flux as the science target, with a relative difference compatible with the error budget on the visibility. A difference of flux between the science and the photometry target of x\% will introduce a relative error of the same x\% on the visibility estimated using Eq.~(\ref{Vstar}). An alternative way is to use a bright photometric calibrator to estimate the photometric bias that is currently of about 0.2~Jy in $L$ and~5~Jy in $N$ with the ATs. This requires calibrators brighter than 1.5~Jy in $L$ and 50~Jy in $N$ for 10\%-accuracy estimates. The photometric calibrator does not need to have a well-defined spatial structure but its flux and spectrum within the field of MATISSE must be known with the above accuracy. We see that the photometric calibration is a major issue in $N$.

	\subsection{Differential visibility and coherent flux ratio}

MATISSE can use many observables that are not sensitive to photometric calibration errors, such as:
\begin{itemize}
\item the differential visibility  in each spectral channel $V_{ij}~/~\overline{V}_{ij(\lambda)}$, where $\overline{V}_{ij(\lambda)}$ is the average visibility over the spectral bandwidth excluding the reference spectral channel;
\item the coherent flux ratio  in each spectral channel $C_{ij}~/~C_{\rm B_{min}}$, where all baselines are calibrated by (for example) the shortest one $B_{\rm min}$.
\end{itemize}
These measurements still need a calibration of the intrumental response in visibility $R_{ij}$. The relevant interferometric calibrator must be a point source or a disk with an accurate diameter. It must be brighter than the science target. In the $L$-band, $R_{ij}$ is quite sensitive to the seeing conditions and mainly to the coherence time $\tau_{0}$. It is therefore important to observe it as close as possible in space and time to the science target. For accurate instrument visibility calibration, it is recommended to use a "Cal1-Sci-Cal2" sequence with an average airmass for the 2 calibrators identical within 3\% to that of the science target. For coherent fluxes larger than about 10~Jy in $L$ and~90~Jy in $N$ with ATs, the contribution of fundamental noise becomes small with regard to the other contributions. In the $N$-band, $R_{ij}$ is much less sensitive to atmospheric variations, and the calibrator should be brighter than the science target, without companion and infrared excess, regardless of its proximity to the science target.

	\subsection{Differential phase}

The differential phase  in each spectral channel $\phi_{ij}~-~\overline{\phi}_{ij(\lambda)}$ (with the same definition of the reference channel than for the differential visibility), representing the change of phase with wavelength, is sensitive only to instrument artifacts (mostly detector features) and the chromatic optical path difference (OPD)  introduced by the atmosphere and the difference of airpath in the delay line tunnels. The detector features are in principle calibrated by an internal MATISSE calibration. At this point, we do not have a reliable and accurate model for the chromatic OPD and it is recommended to calibrate the differential phase with calibrators introducing the same air path difference, i.e. at the same declination and with hour angles allowing them to be observed in the same position on the sky than the science target, with calibrators brighter than their science target.

	\subsection{Closure phase}
The closure phase  in each spectral channel  $\psi_{ijk}~=~\phi_{ij}~+~\phi_{jk}+~\phi_{ki}$ is in principle self-calibrated by MATISSE, but it remains sensitive to fast detector variations and it can be slightly contaminated by the chromatic OPD that affects the differential phase, for example because of crosstalk between MATISSE beams or fringe peaks. It is therefore recommended to calibrate it with the same calibrators as for the differential phase.

	\subsection{Imaging runs}
During imaging runs, many observations of a science target are repeated. They should be merged with calibrators fulfilling all the conditions above:
\begin{itemize}
\item Cal1L and Cal2L for instrument and atmosphere calibration in the $L$-band. At least one of them should be choosen to be also a good chromatic OPD calibrator.
\item CalPL for a photometric calibration in the $L$-band.
\item CalPN for photometric calibration in the $N$-band.
\item Cal3N for interferometric calibration in the $N$-band.
\end{itemize}
The photometric calibrator can be used only once per sequence, while the interferometric ones must be repeated regularly. Of course, one should try to reduce the number of calibrators choosing for instance Cal2L=CalPL and Cal3N=CalPN.

\section{The primary lists of calibrator candidates for VLTI/MATISSE } \label{section:appli}

   \begin{figure}
   \centering
  \includegraphics[trim = 0cm 0cm 0cm 0cm, clip, width=\hsize]{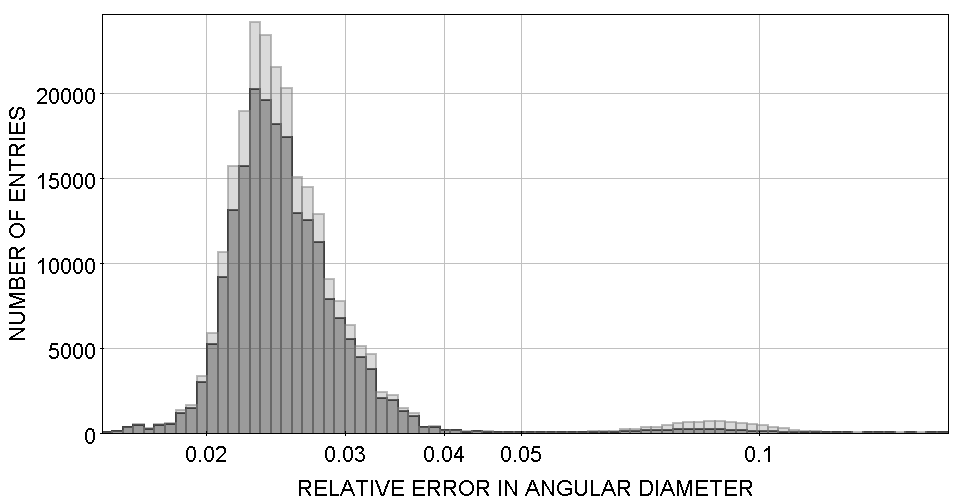}
      \caption{Distribution of the relative error in the LDD diameter reported in the JSDC for the "pure" (dark grey) and excess-free (grey) calibrators. 
              }
         \label{fig:RelError}
   \end{figure}

   \begin{figure}
   \centering
  \includegraphics[trim = 0cm 0cm 0cm 0cm, clip, width=\hsize]{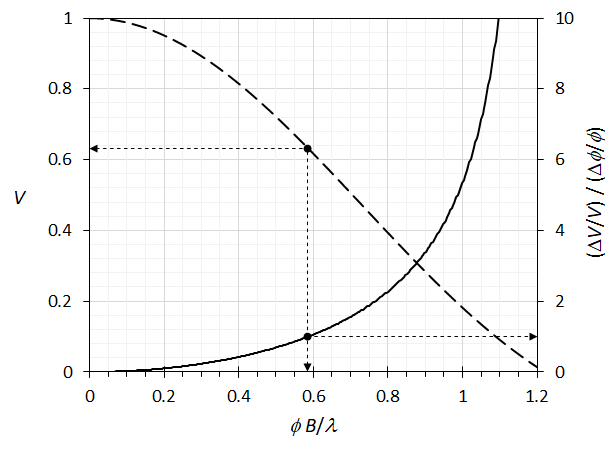}
      \caption{Ratio of the relative error in UD visibility to the relative error in angular diameter (solid line, right scale), and UD visibility (dashed line, left scale), versus the angular diameter in angular resolution unit. The black dots are the points of the 2 curves of same abscissa $\phi~B/\lambda~\sim~0.59$ (corresponding to $\Delta V/V~=~\Delta \phi/{\phi}$).
              }
         \label{fig:Error}
   \end{figure}

   \begin{figure}
   \centering
  \includegraphics[trim = 0cm 0cm 0cm 0cm, clip, width=\hsize]{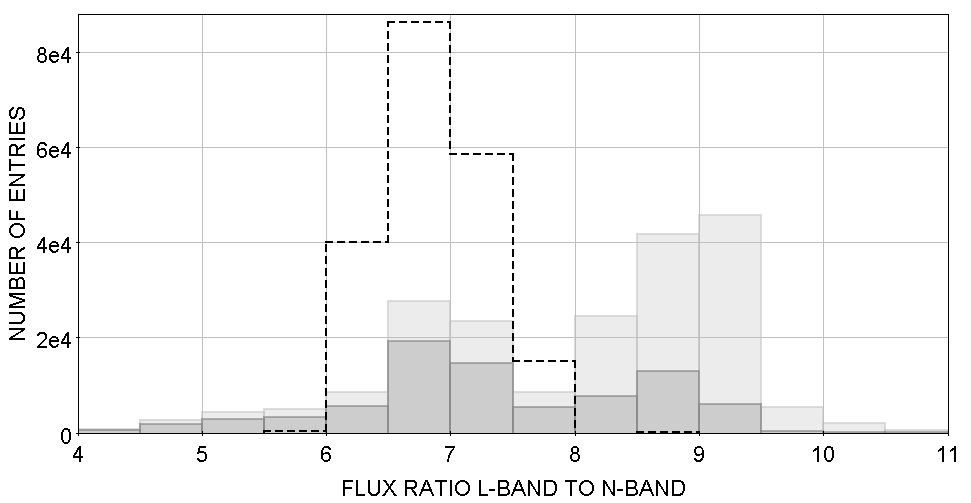}
      \caption{Distribution of the ratio of the flux density in $L$ to $N$ for the "pure" calibrators (light grey: all spectral types; dark grey: late-type stars). The dashed histogram corresponds to the flux ratio $L$ to $N$ for the pure calibrators of all spectral types assumed to be blackbodies. 
              }
         \label{fig:FlxRatio}
   \end{figure}

%
\begin{table*}
\caption{Number of entries for each brightness-limit subset of calibrator candidates suitable with ATs and UTs in the $L-$ and $N-$bands for different values of IRflag. "Hybrid" calibrators are for both bands. "Med+" stands for "Medium-bright+bright" and "Faint+" stands for "Faint+Medium-bright+bright". The "bright", "medium-bright", and "faint" categories are defined in Table~\ref{table:categories}. The 4 numbers in bold are those of the 4 final primary lists of excess-free calibrator candidates.}             
\label{table:summary}      
\centering                          
\begin{tabular}{l  l | c c c | c c c }        
\hline\hline                 
  & & \multicolumn{3}{c |} {ATs}  & \multicolumn{3}{c} {UTs} \\
 Subset  & IRflag &  \multicolumn{1}{c }{$L$}  & \multicolumn{1}{c}{$N$} & \multicolumn{1}{c |}{Hybrid} &  \multicolumn{1}{c}{$L$}  &   \multicolumn{1}{c }{$N$}  & \multicolumn{1}{c }{Hybrid} \\
\hline                        
Bright &  "0" to "7" & 2\,851  & 17  & 1   & 61\,500 & 1\,295  & 455 \\
 &  "0", "2", "4", or "6" & \textbf{1\,621}  & 3 &  0  & 56\,068 & 581 & \textbf{259}\\
 & "0" & 171  & 0 &  0  & 41\,269 & 28 & 14\\
 \hline   
Med+ & "0" to "7" &  29\,991  & 142   & 3  &  201\,468  & 5\,293  & 2\,621  \\
 &  "0", "2", "4", or "6" &  26\,314 & \textbf{44}  & 0 & 158\,715 & 2\,412 &  1\,476\\
 & "0" &  16\,886 & 1  & 0 & 132\,861 & 237 &  156\\
 \hline   
Faint+ &  "0" to "7"  &  142\,620 & 1\,642 & 656  & 300\,332  & 49\,345  & 40\,077  \\
 & "0", "2", "4", or "6"  & 122\,958 & 750 & \textbf{375} & 171\,663 & 41\,288 & 36\,116 \\
 & "0"  & 101\,295 & 40 & 22 & 144\,113  & 27\,751 & 24\,828 \\
\hline\end{tabular}
\end{table*}

	\subsection{Requirements}

From our catalogue, we extract lists of calibrator candidates for the $L$- and $N$-bands, based on their angular diameter, object type, degree of binarity, infrared brightness and features. The selected targets are bright point-like stars observable with the VLTI/MATISSE. Since our goal is to provide a self-consistent interferometric network of calibrators suitable for VLTI/MATISSE, we have started a large observing programme in order to measure their angular diameter with  high accuracy (1\% or even better) in the MIR spectral bands.

	\subsection{Selection criteria} \label{criteria}

The candidate calibration stars must satisfy the following
conditions, listed in the order they are applied to the building process of our catalogue:

\begin{enumerate}

\item The source must be observable from the ESO-Paranal Observatory (latitude 24\degr 40' S). This condition provides a subset of 371\,333 candidates.

\item The source must be a potential calibrator suitable for optical  long-baseline interferometry (CalFlag of  "0"). This flag condition provides a subset of 359\,580 calibrator candidates observable from Paranal.

\item The source must be as small as possible in order to minimize the calibration error caused by the uncertainty in the calibrator modeling. Figure~\ref{fig:RelError} shows the distribution of the relative error in the LDD diameter given by the JSDC for the "pure" (IRflag of  "0") and excess-free (IRflag of  "0",  "2",  "4", or  "6") calibrators. The mean value of the relative error in angular diameter is about 2.6-2.7\%. Using the UD model, one can  demonstrate \citep[see e.g.][]{Borde02} that the relative error in visibility $\Delta V/V$ remains lower than the relative error in angular diameter $\Delta \phi / \phi$ provided that $\phi~<~0.59~\lambda/B$  (corresponding to $V$~>~0.63), as shown in Fig.~\ref{fig:Error}. For the 130-m projected baselength this condition corresponds to $\phi$~<~3.3~mas for $\lambda$~=~3.5~$\mu$m, and $\phi$~<~10~mas for $\lambda$~=~10.5~$\mu$m. Choosing targets with $\phi$~<~3~mas for the $L$-band or $\phi$~<~9~mas forthe $N$-band ensures that $\Delta V / V$~<~0.8~$\Delta \phi / \phi $, in both bands for any baseline smaller than 130~m ($\phi~B/\lambda$~<~0.54; $V$~>~0.68). With this size condition, 15 sources are excluded from the subset for the $N$-band, while 542 sources are excluded for the $L$-band.

\item To ensure a good confidence level in the final flux estimate, we exclude the sources  with a large dispersion of the photometric points and those with less than 2 photometric points. Since 90\% of the entries of our catalogue have $rdisp$~<~0.14 in $L$ or $rdisp$~<~0.29 in $N$ ($rdisp$ is the relative dispersion, given by the ratio of the dispersion to the flux value), we exclude the sources with $rdisp$~>~0.15  in $L$ and $rdisp$~>~0.3 in $N$. Observing with the VLTI/MATISSE on all four UTs is a rather expensive undertaking and it is therefore important to reduce the on-sky calibration time as much as possible. Therefore we also look for calibrators that are both compact enough to serve as $L$-band calibrators ($\phi$~<~3~mas), while still being bright enough in the $N$-band to also serve as $N$-band calibrators. We call them "hybrid" calibrators. Using the ATs, 304\,816 candidates are kept for $L$, 307\,815 candidates for $N$, and 285\,688 are hybrid calibrators.

\item The sources must be detectable with MATISSE and useable at least for coherent flux measurements. Thus, they must belong to the category of "faint" sources or even brighter, as defined  in Table~\ref{table:categories}:
\begin{itemize}
\item With ATs, this corresponds to $C_L$~>~0.2~Jy and  $C_N$~>~20~Jy. We find 142\,620 potential candidates suitable for $L$, 1\,642 for $N$, and 656 for both bands (hybrid).
\item With UTs, this corresponds to $C_L$~>~0.015~Jy and $C_N$~>~0.15~Jy. We find 300\,332 potential candidates suitable for $L$, 49\,345 for $N$, and 40\,077 for both bands (hybrid).
\end{itemize}

\end{enumerate}

        \subsection{Results}

Table~\ref{table:summary} gives the final numbers of $L$-band, $N$-band, and  hybrid candidates suitable with the ATs and the UTs, for each subset of brightness limit (bright targets, "med+"=bright+medium-bright targets, "faint+''=bright+medium-bright+faint targets), and for different values of IRflag. 

We find only 1 hybrid bright "pure" calibrator candidate suitable for the ATs  (V343~Pup), and 14 "pure" calibrator candidates suitable for the UTs. 
To explain the small numbers of hybrid "pure" calibrators, we invoke the fact that the "standard" stars, i.e. showing no IR-excess, have their flux density in $L$ always higher than in $N$. The statistics of the flux ratio $L$ to $N$ for the  "pure" calibrators is:
\begin{itemize}
\item lower quartile ratio $F_L$/$F_N$:~7;
\item median ratio:~8.4;
\item upper quartile ratio:~9. 
\end{itemize} 
Figure~\ref{fig:FlxRatio} shows that the distribution of the flux ratio for the "pure" calibrators reveals 2 peaks around the values of 7 and 9. If we compute the flux ratio at $\lambda~=~3.5~\mu{\rm m}$ to the flux at $\lambda~=~10.5~\mu{\rm m}$ for the same stars assumed to be blackbodies, we find that the distribution of the flux ratio shows a single peak around 7 (dashed overplotted histogram of Fig.~\ref{fig:FlxRatio}). We suspect the second peak of the flux ratio distribution (around 9) to be caused by stars with SEDs deviating from simple blackbodies, showing sort of "infrared deficits" probably caused by the presence of absorption bands in this spectral domain. 

As primary lists of calibrator candidates, we select the lists of  excess-free candidates (IRFlag of  "0",  "2",  "4", or  "6"): hybrid-bright for the UTs (259 entries); $L$-bright (1\,621 entries), $N$-"med+" (44 entries), and hybrid-"faint+" (375 entries) for the ATs.
We note that all the 259 hybrid UT-bright candidates are $L$-band AT-bright candidates, and are hybrid AT-"faint+" candidates as well. All the 375 hybrid AT-"faint+" candidates are also $L$-band AT-bright candidates.

Matching our primary lists with the MIDI and the Cohen lists formely used for MIR interferometry, we find 95 $L$-band AT-bright candidates that are MIDI calibrators and 127 that are Cohen's standards; 22 $N$-band AT-"med+" candidates that are MIDI calibrators and 18 that are Cohen's standards; 79 hybrid AT-"faint+"  candidates that are MIDI calibrators and 113 that are Cohen's standards;  79 hybrid UT-bright candidates that are MIDI calibrators and 103 that are Cohen's standards. A detailed analysis of the original MIDI and Cohen's lists reveals that only 10 MIDI calibrators and 15 Cohen's standards are "pure" calibrators (with null CalFlag and IRflag), which may suggest that our flag criteria to identify the "pure" calibrators are much more selective than the criteria used to build the MIDI and the Cohen's lists. Relaxing the IRflag constraint, we find that 244 MIDI calibrators (among 402) and 291 Cohen's standards (among 422) are excess-free calibrators (with null CalFlag and IRflag of "0", "2", "4", or "6"). 

Table~\ref{table:AT-hybrid} reports the first 10 rows of the list of the 375 hybrid AT-"faint+" excess-free calibrator candidates, ordered by their increasing right ascension. For brevity, we give the name of each selected source, with its spectral type, coordinates (J2000), angular diameter values with associated errors reported in the MIDI list, the Cohen's list, and the JSDC, flux and correlated flux values with errors in the $L$- and $N$-bands. To get a rough estimate of the relative uncertainty in correlated flux, we simply add the relative uncertainty in angular diameter to the relative dispersion in flux. The value of IRflag is also reported in the list. The full table (containing the 375 entries) is available online.

%

%
\begin{landscape}
\begin{table}
\caption{Sample table showing the first 10 hybrid AT-"faint+" excess-free calibrator candidates, ordered by their increasing right ascension. $\phi_{\rm MIDI}$, $\phi_{\rm Cohen}$, and $\phi_{\rm JSDC}$ are the angular diameter estimates, respectively reported in the MIDI list, the Cohen's list, and the JSDC.  $\Delta\phi_{\rm MIDI}$,  $\Delta\phi_{\rm Cohen}$, and $\Delta\phi_{\rm JSDC}$ are their associated uncertainties. $F_{L}$ and $F_{N}$ are the flux density in bands $L$ and $N$ respectively. $\Delta F_{L}$ and  $\Delta F_{N}$ are their associated uncertainties. $C_{L}$ and $C_{N}$ are the correlated flux for the UD model with the 130-m baselength in the $L$- and $N$-bands respectively. $\Delta C_{L}$ and  $\Delta C_{N}$ are their associated uncertainties. The full table (containing the 375 entries) is available online.}
\label{table:AT-hybrid} 
\begin{tabular}{l l c c c c c c c c c c c c c c c c c  }
\hline
Name &  Spectral & $\alpha$J2000 & $\delta$J2000 & $\phi_{\rm MIDI}$ & $\Delta\phi_{\rm MIDI}$ & $\phi_{\rm Cohen}$ & $\Delta\phi_{\rm Cohen}$ & $\phi_{\rm JSDC}$ & $\Delta\phi_{\rm JSDC}$ & IRflag  & $F_{L}$ &  $\Delta F_{L}$  & $F_{N}$ &  $\Delta F_{N}$  & $C_{L}$ &  $\Delta C_{L}$  & $C_{N}$ &  $\Delta C_{N}$ \\
 & Type & (h:m:s) & (d:m:s) & (mas)  & (mas) & (mas) & (mas) & (mas) & (mas) & &  (Jy) &  (Jy) & (Jy) &  (Jy) &  (Jy) &  (Jy) & (Jy) &  (Jy)  \\
\hline
kap02 Scl & K2III & 00:11:34.419 & -27:47:59.03 & - & - & 1.73 & 0.03 & 1.72 & 0.15 & 2 & 34.7 & 4.7 & 5.6 & 1.4 & 30.9 & 7.0 & 5.6 & 1.8\\ 
HD 853 & M2III & 00:12:55.139 & -03:22:53.56 & - & - & - & - & 2.25 & 0.16 & 6 & 18.4& 1.9 & 5.0 & 1.2 & 15.0 & 2.7 & 4.9 & 1.5\\ 
HD 942 & K5III & 00:13:42.245 & -26:01:20.44 & - & - & - & - & 2.26 & 0.19 & 6 & 43.1 & 4.8 & 6.9 & 0.9  & 35.1 & 6.9 & 6.8 & 1.5\\ 
HD 1063 & M0III & 00:14:55.762 & -03:01:38.69 & - & - & - & - & 1.97 & 0.16 & 6 &24.0  & 1.5 & 4.3 & 1.2  & 20.6 & 3.0 & 4.2 & 1.6\\ 
HD 1187 & K2III & 00:16:08.867 & -31:26:47.02 & - & - & - & - & 1.56 & 0.15 & 6 & 27.0 & 1.0 & 4.1 & 1.0 & 24.5 & 3.3 & 4.0 & 1.4\\ 
HD 1923 & M2III & 00:23:22.145 & -29:50:50.11 & - & - & - & - & 2.42 & 0.19 & 0 & 39.0 & 2.5 & 5.6 & 1.3 & 30.7 & 4.4 & 5.5 & 1.7\\ 
eps And & G7III & 00:38:33.346 & +29:18:42.31 & - & - & - & - & 1.86 & 0.16 & 6 & 40.2 & 1.9 &  6.4 & 1.5  & 35.0 & 4.6 & 6.3 & 2.0\\ 
mu Phe & G8III & 00:41:19.552 & -46:05:06.02 & - & - & - & - & 1.75 & 0.15 & 6 & 32.6 & 2.9 & 5.2 &  0.4 & 28.9 & 5.0 & 5.2 & 0.8\\ 
CI Psc & M3III & 00:44:55.279 & +03:12:03.59 & - & - & - & - & 2.30 & 0.21 & 2 & 32.7 & 1.8 & 5.7 &  1.7 & 26.3 & 3.8 & 5.6 & 2.2\\ 
HD 5098 & K1III & 00:52:40.625 & -24:00:21.04 & - & - & - & - & 1.60 & 0.16 & 6 & 28.5 & 0.2 & 4.5 & 1.1 & 25.7 & 2.7 & 4.4 & 1.5\\ 
... &  &   &   &  &  &  &  &   &   &   &   &   &   & & & & &    \\ 
\hline
\end{tabular}
\end{table}
\end{landscape}

        \subsection{Statistics of the primary lists}

   \begin{figure*}
   \centering
  \includegraphics[trim = 0cm 0cm 0cm 0cm, clip, width=15cm]{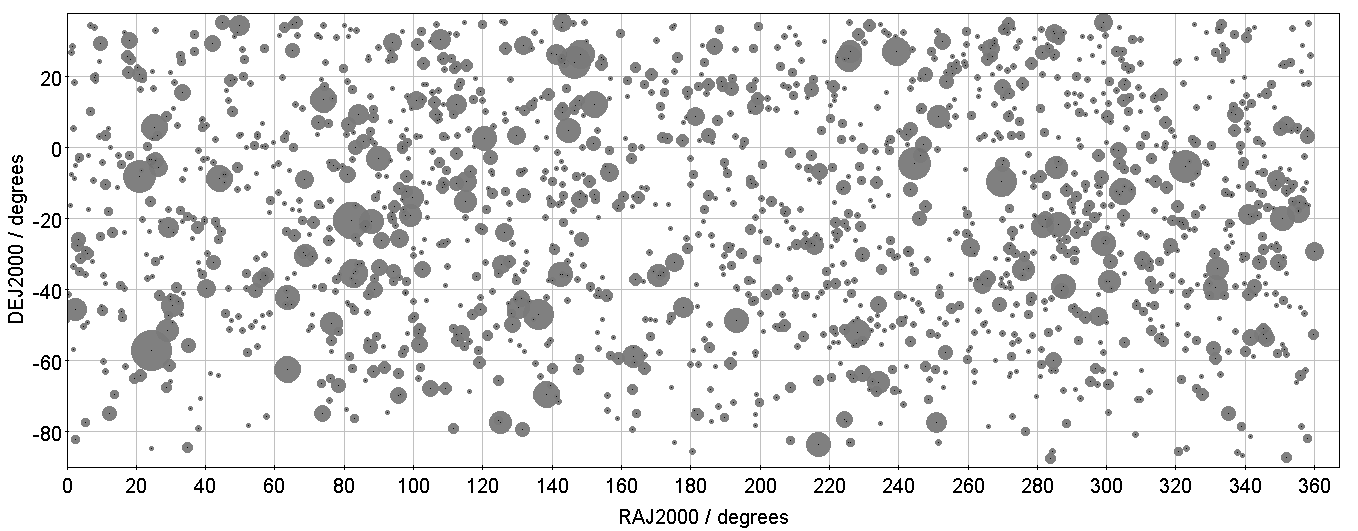}
      \caption{Sky coverage of the 1\,621 $L$-band AT-bright excess-free calibrator candidates. The size of the dots is proportional to $C_{L}$.}
         \label{fig:SkyCovCalLAT}
   \end{figure*}

   \begin{figure*}
   \centering
  \includegraphics[trim = 0cm 0cm 0cm 0cm, clip, width=15cm]{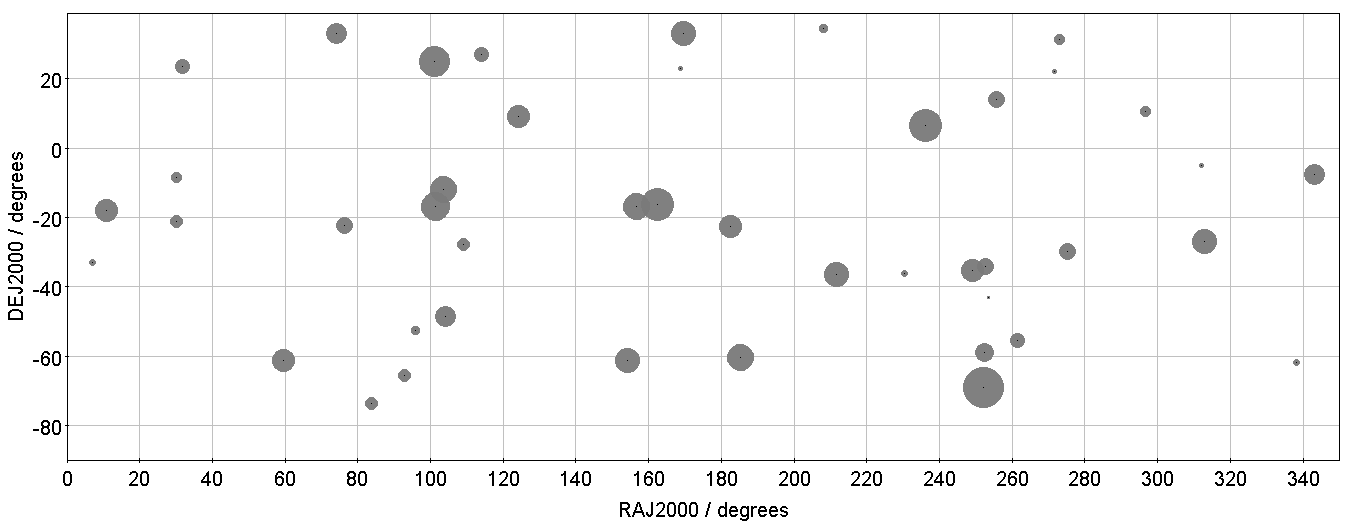}
      \caption{Sky coverage of the 44 $N$-band AT-"med+" calibrator candidates. The size of the dots is proportional to $C_{ N}$.}
         \label{fig:SkyCovCalNAT}
   \end{figure*}

   \begin{figure*}
   \centering
  \includegraphics[trim = 0cm 0cm 0cm 0cm, clip, width=15cm]{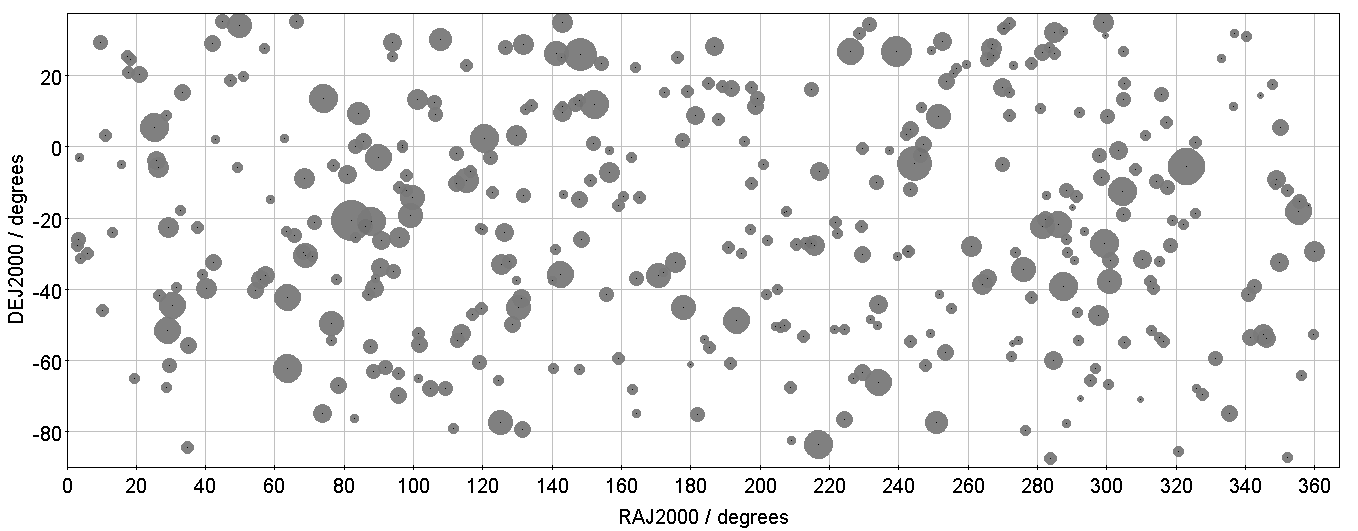}
      \caption{Sky coverage of the 375 hybrid AT-"faint+" calibrator candidates. The size of the dots is proportional to $C_{ N}$.}
         \label{fig:SkyCovCalHybridAT}
   \end{figure*}

   \begin{figure*}
   \centering
  \includegraphics[trim = 0cm 0cm 0cm 0cm, clip, width=15cm]{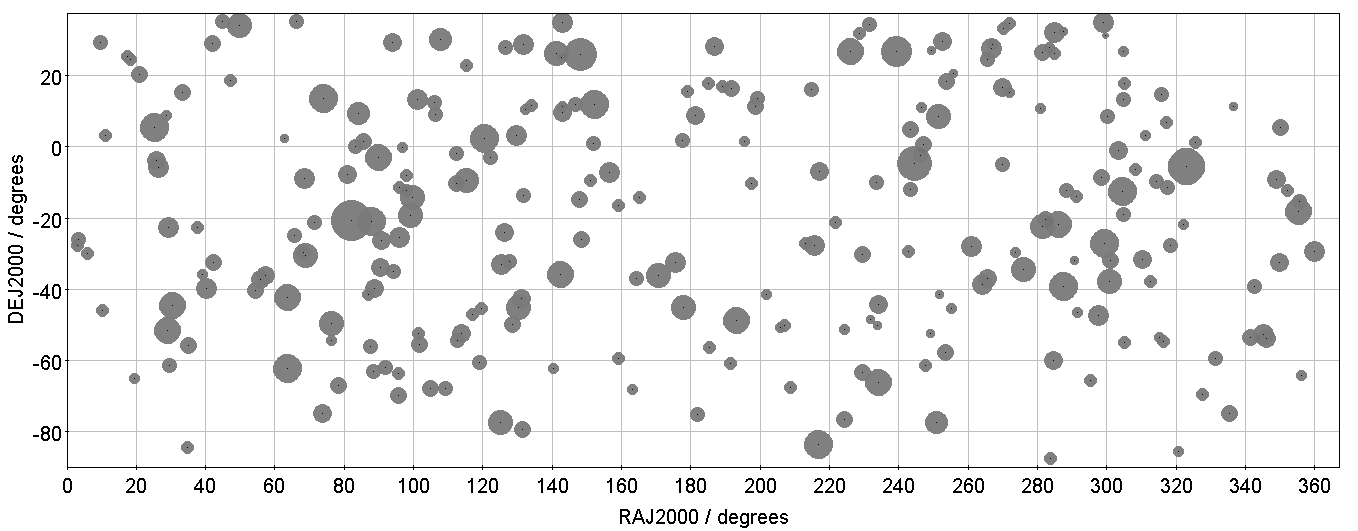}
      \caption{Sky coverage of the 259 hybrid UT-bright excess-free calibrator candidates. The size of the dots is proportional to $C_{ N}$.}
         \label{fig:SkyCovCalHybridUT}
   \end{figure*}

   \begin{figure}
   \centering
  \includegraphics[trim = 0cm 0cm 0cm 0cm, clip, width=\hsize]{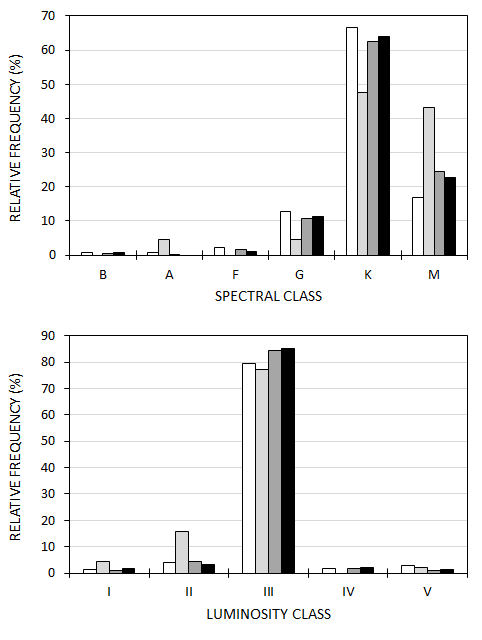}
      \caption{Distribution in spectral (top) and luminosity (bottom) class of the $L$-band AT-bright (white), $N$-band AT-"med+" (light grey), hybrid AT-"faint+" (dark grey), and hybrid UT-bright (black) excess-free calibrator candidates.
              }
         \label{fig:SpTypeCal}
   \end{figure}

   \begin{figure}
   \centering
  \includegraphics[trim = 0cm 0cm 0cm 0cm, clip, width=\hsize]{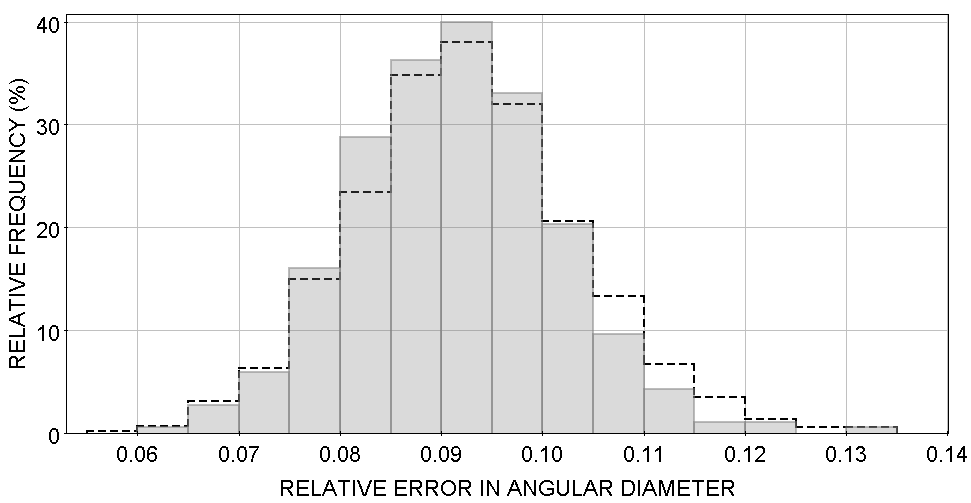}
      \caption{Distribution of the relative error in angular diameter reported in the JSDC for the 1\,621 $L$-band AT-bright (black thin dashed steps) and the 375 hybrid AT-"faint+" (light grey filled bars) excess-free calibrator candidates. 
              }
         \label{fig:RelErrorCal}
   \end{figure}

Figures~\ref{fig:SkyCovCalLAT} to \ref{fig:SkyCovCalHybridUT}  respectively show the sky coverage of the $L$-band AT-bright, $N$-band AT-"med+", hybrid AT-"faint+", and hybrid UT-bright  excess-free calibrator candidates. In each figure, the size of the dots  is proportional to the 130-m UD correlated flux: in the $L$-band for Fig.~\ref{fig:SkyCovCalLAT}, in the $N$-band for the 3 other figures. 

From our study, we conclude that:
\begin{enumerate}
\item Finding bright calibrators closer than a few degrees to any scientific target is not an issue for the $L$-band, with both the UTs and the ATs. 
\item With UTs, finding bright excess-free calibrator candidates suitable for both the $L$- and the $N$-bands (hybrid) is guaranteed in a circle of about 10-15\degr~around any scientific target.
\item With ATs, no source of the calalogue can be used as a bright calibrator candidate suitable for both bands. To find an hybrid calibrator in a circle of 10\degr~around any scientific taget, we need to include the medium-bright and the faint sources ("faint+").
\item With ATs in the $N$-band, our catalogue gives only 3 bright and 44 bright+medium-bright ("med+") excess-free calibrator candidates observable with the VLTI/MATISSE. Unless including the faint sources as well, the average distance from target to  "med+" excess-free calibrator is rarely less than 20-25\degr.
\end{enumerate}

Figure~\ref{fig:SpTypeCal} shows the distribution in spectral and luminosity classes of the calibrators for each list. We count 1\,290 K-giants (with luminosity class III) in the $L$-band AT-bright list; 34 in the $N$-band AT-"med+" list; 316 in the hybrid AT-"faint+" list; and 221 in the hybrid UT-bright list. 

Figure~\ref{fig:RelErrorCal} shows the distribution of the relative error in angular diameter reported in the JSDC for the $L$-band AT-bright and the hybrid AT-"faint+" excess-free calibrator candidates. The mean relative error in angular diameter is 10\% for the $N$-band AT-"med+" excess-free calibrator candidates and 9\% for the excess-free calibrator candidates of the 3 other lists. These values are significantly higher than the mean value of 2.6\% obtained with the complete initial set of calibrator candidates (see Fig.~\ref{fig:RelError}). As shown in Fig.~\ref{fig:Error}, to obtain visibility measurement within 1\% accuracy, a calibration error of 10\% level needs the use of UD calibrators with $\phi$~<~ 1.1~mas for the $L$-band, and $\phi$~<~3.3~mas for the $N$-band with the 130-m baselength  ($\phi~B/\lambda$~<~0.2; $V_{\rm ud}$~>~0.95). 

Since \citet{chelli16} claimed that the method they used to estimate the angular diameter reported in the JSDC (derived from the surface brightness method) does not depend on the luminosity class, we invoke the spectral class only to explain this discrepancy in diameter uncertainty. In our lists, the K and M stars represent 83\% of the $L$-band AT-bright, 91\% of the $N$-band AT-"med+", and 87\% of both the AT-"faint+" and the UT-bright hybrid  excess-free calibrator candidates, while they represent only 40\% of the total set of excess-free calibrator candidates in our catalog. The high level of diameter uncertainty (9-10\%) reported in the JSDC for our calibrator candidates confirms the crucial need for measuring the angular diameter of the calibrators for the VLTI/MATISSE (mainly giant stars) with 1\% accuracy (or even better).

\section{Conclusions}

We have built a new all-sky catalogue, called the Mid-infrared stellar Diameter and Flux compilation Catalogue (MDFC), that contains 465\,857 entries with the aim of helping the users of long-baseline interferometers operating in the mid-infrared with the preparation of their observations and the calibration of their measurements. The main improvement to the other existing catalogues  is the specific extension to the mid-infrared wavelengths. Our catalogue covers the entire sky and contains mainly dwarf and giant stars from A to K spectral types, closer than 6~kpc. The smallest values of the reported median value in flux density are 0.16~mJy in the $L$-band and 0.1~mJy in the $N$-band. The smallest values of the reported diameter reach 1~microarcsec. The construction of the catalogue is divided in 3 main steps:
\begin{enumerate}
\item The angular diameter estimates reported in the JSDC are complemented by the measurements compiled in the JMDC, the diameter estimates computed from the distance and the radius reported in the \textit{Gaia} DR2 data release, and the diameter estimates reported in the VLTI/MIDI list of calibrator candidates and in the Cohen's list of spectrophotometric standards.
\item The median flux density and an estimate of the statistical dispersion in the $L$-, $M$-, and $N$-bands are computed from the compilation of almost 20 photometric catalogues. 
\item The information about the presence of circumstellar features around each source revealed by the IR data (excess, extent, and variability) is synthetized into a single 3-bit flag.
\end{enumerate}

Our infrared flag, which is used complementary to the calibrator flag reported in the JSDC, allows us to report a list of  201\,200 "pure" calibrators suitable for the mid-infrared. These sources are single stars or wide binaries with a favorable object type, have a reliable angular diameter estimate, and show no evidence for IR feature (excess, extent, variability). 

Selecting only the excess-free calibrators, we produce the 4 primary lists of VLTI/MATISSE calibrators containing more than one thousand of them (mainly cool giants): with ATs for the $L$-band (1\,621 bright sources), for the $N$-band (44 bright and medium-bright sources), and for both bands (375 hybrid bright, medium-bright, and  faint sources); with UTs for both bands (259 hybrid bright sources). They are selected according to:
\begin{enumerate}
\item  their declination;
\item the reliability and value of their angular diameter estimate;
\item their astrometric binarity; 
\item their SIMBAD Object Type;
\item the reliability and value of their MIR flux and correlated flux estimates; and
\item their IR excess. 
 \end{enumerate}
 Since they have not yet been measured or modeled with a better accuracy than 5\%, we have initiated a large observing programme with the VLTI/MATISSE in the aim to measure their angular diameter at 1\% accuracy. To get this challenging accuracy without any need of external calibration, we have developed a new method that will be the subject of a forthcoming paper.

\section*{Acknowledgements}

    This work uses the VizieR catalogue access tool,
    CDS, Strasbourg, France; the SIMBAD database, operated at 
    CDS, Strasbourg, France; and the TOPCAT software, provided by
    Mark Taylor of Bristol University, England available at
    starlink.ac.uk/topcat. \\
This publication makes use of the data products from:\\
- The \textit{Hipparcos} and Tycho Catalogues; \\
- The Tycho-2 Catalogue; \\
- The \textit{Gaia} data release 2 (DR2), \textit{Gaia} team;\\
- The Estimating distances from \textit{Gaia} DR2 parallaxes catalog, Coryn Bailer-Jones calj@mpia.de;\\
- The $UBVRIJKLMNH$ Photoelectric Photometric Catalogue;\\
- The Catalogue of 10-micron celestial objects;\\
- The IRAS Catalog of Point Sources, Version 2.0, Joint IRAS Science W.G.;\\
- The IRAS Faint Source Catalog, |b| > 10, Version 2.0;\\
- The Two Micron All-Sky Survey, which is a joint project of the University of Massachusetts and the Infrared Processing and Analysis Center/California Institute of Technology, funded by the National Aeronautics and Space Administration and the National Science Foundation;\\
- The Galactic Legacy Infrared Mid-Plane Survey Extraordinaire (GLIMPSE), Spitzer Science Center;\\
- The AKARI/IRC Mid-Infrared All-Sky Survey (Version 1);\\
- The Wide-field Infrared Survey Explorer, which is a joint project of the University of California, Los Angeles, and the Jet Propulsion Laboratory/California Institute of Technology, funded by the National Aeronautics and Space Administration, Roc Cutri (IPAC/Caltech);\\
- The IRAS PSC/FSC Combined Catalogue;\\
- The JMDC: JMMC Measured Stellar Diameters Catalogue, Gilles Duvert gilles.duvert@univ-grenoble-alpes.fr;\\
- The Jean-Marie Mariotti Center JSDC catalog, downloadable at http://www.jmmc.fr/jsdc, Gilles Duvert gilles.duvert@univ-grenoble-alpes.fr;\\
- The MSX6C Infrared Point Source Catalog;\\
- The Washington Double Star Catalog maintained at the U.S. Naval Observatory;\\
- The COBE DIRBE Point Source Catalog, Beverly J. Smith beverly@nebula.etsu.edu;\\
- The Parameters and IR excesses of Gaia DR1 stars catalogue, Iain McDonald iain.mcdonald-2@manchester.ac.uk;\\
- The Stars with calibrated spectral templates list, downloadable at http://www.gemini.edu/sciops/instruments/mir/Cohen\_list.html;\\
- The VLTI/MIDI calibrator candidates list, downloadable at http://ster.kuleuven.be/$\sim$tijl/MIDI\_calibration/mcc.txt.\\
\\
P.C.  wants  to  thank A.~Chiavassa, O.~Creevey,  D.~Mourard, N.~Nardetto,  M.~Schultheis,  and F.~Th\'{e}venin for their helpful comments and advice.

\bibliographystyle{mnras}
\bibliography{bidfile}

\appendix

\section{Used catalogues} \label{Catalogs}

\begin{landscape}
\begin{table*}
\caption{Catalogues and lists used to build the MDFC catalogue.}\label{table:catalogs}     
\begin{tabular}{ l l r l }     
\hline    
Catalogue/list & Title &\# of entries & Ref. \\ 
\hline                    
          SIMBAD & The SIMBAD/CDS database & 8\,230\,006 & \citet{wenger00} \\
           I/239/tyc\_main & The \textit{Hipparcos} and Tycho Catalogues &  1\,058\,332 & \citet{esa97} \\  
           I/345/gaia2 & \textit{Gaia} data release 2 (DR2) &  1\,692\,919\,135 & \citet{gaia18}  \\
           I/347/gaia2dis & Estimating distances from \textit{Gaia} DR2 parallaxes & 1\,331\,909\,727 & \citet{bailerjones18} \\
           II/7A/catalog & UBVRIJKLMNH Photoelectric Catalogue & 5\,943 & \citet{morel78} \\
           II/53/catalog &  Catalogue of 10-micron Celestial Objects & 647 & \citet{hall74} \\
         II/125/main & IRAS catalogue of Point Sources, Version 2.0  & 245\,889 & \citet{helou88} \\
         II/156A/main & IRAS Faint Source Catalogue, |b| > 10, Version 2.0 & 173\,044 & \citet{moshir90} \\
          II/246/out & 2MASS All-Sky Catalogue of Point Sources & 470\,992\,970 & \citet{cutri03} \\ 
          II/293/glimpse & GLIMPSE Source Catalogue (I + II + 3D) & 104\,240\,613 & \citet{benjamin03} \\
          II/297/irc & AKARI/IRC mid-IR all-sky Survey  & 870\,973 & \citet{ishihara10} \\
          II/311/wise & WISE All-Sky Data Release & 563\,921\,584 & \citet{wright10} \\
          II/328/allwise & AllWISE Data Release & 747\,634\,026 & idem \\
          II/338/catalog & IRAS PSC/FSC Combined Catalogue & 345\,162 & \citet{abrahamyan15} \\
          II/345/jmdc & JMMC Measured Stellar Diameters Catalogue & 1\,554 &\citet{duvert16} \\ 
          II/346/jsdc\_v2 & JMMC Stellar Diameters Catalogue & 465\,877 & \citet{bourges17} \\
           V/114/msx6\_gp &  The complete MSX6C catalogue in the Galactic Plane (|b|$\le$6\degr) & 431\,711 &\citet{egan03} \\
           V/114/msx6\_main &  MSX6C Infrared Point Source Catalogue, High latitude, |b| > 6 & 10\,168 & idem \\
           B/wds/wds & Washington Double Star Catalogue & 142\,596 & \citet{mason01}\\ 
  J/ApJS/154/673/DIRBE & COBE DIRBE Point Source Catalogue & 11\,788 & \citet{smith04}\\
 J/MNRAS/471/770/table1 & Param. and IR exc. of \textit{Gaia} DR1 stars. Tycho-Gaia astrom. solution & 1\,475\,921 & \citet{mcdonald17} \\
 J/MNRAS/471/770/table2 & Param. and IR exc. of \textit{Gaia} DR1 stars. Hipparcos-Gaia astrom. solution &107\,110 & idem\\
 J/MNRAS/471/770/table3 & Param. and IR exc. of \textit{Gaia} DR1 stars. Stars with IR excess & 4\,255 & idem\\
\hline
mcc.txt$^{a}$ & VLTI/MIDI list of calibrator candidates &  403 & \citet{verhoelst05}\\
980440.tb400.txt$^{b}$  & Stars with calibrated spectral templates & 422 & \citet{cohen99}, Table 4 \\
Cohen\_list.html$^{c}$ & Complete set of Cohen's standards &  435 & Gemini Obs. (MIR Resources)\\
\hline                  
\end{tabular}
\\
\begin{flushleft}
{
$^{a}${http://ster.kuleuven.be/$\sim$tijl/MIDI\_calibration/mcc.txt}\\
$^{b}${https://iopscience.iop.org/article/10.1086/300813/fulltext/980440.tb400.txt}\\
$^{c}${www.gemini.edu/sciops/instruments/mir/Cohen\_list.html}
}
\end{flushleft}
\end{table*}
\end{landscape}

Table~\ref{table:catalogs} lists all the catalogues used to build the MDFC catalogue, with some of their general features: VizieR reference, title, number of entries, and main reference.

\section{Table columns} \label{append:TabCol}

%
\begin{table*}
\caption{Description of the catalogue fields.
}             
\label{table:columns}      
\begin{tabular}{l c l  }        
\hline
Label & Units & Explanations \\ 
\hline                    
Name	&   & SIMBAD main identifier \\
SpType	&  	& SIMBAD spectral type	 \\
RAJ2000	&"h:m:s"	& Barycentric right ascension (ICRS) at Ep=2000.0	\\
DEJ2000	& "d:m:s"	& Barycentric declination (ICRS) at Ep=2000.0\\
distance	& pc	&  Estimated distance from \citet{bailerjones18} \\
teff\_midi	& K	&  Estimate of effective temperature from VLTI/ MIDI	 \\
teff\_gaia	& K	&  Estimate of effective temperature$^{a}$ from Gaia DR2 \citep[from Apsis-Priam, see][]{Andrae18} \\
Comp	&  	&  Components when more than 2 from WDS	\\
mean\_sep	& \arcsec	&  Mean separation from WDS	\\
mag1	& mag	&  Magnitude of First Component from WDS	\\
mag2	& mag	&  Magnitude of Second Component from WDS\\
diam\_midi	& mas	&  Fitted angular diameter from VLTI/MIDI	\\
e\_diam\_midi	& mas	&  Error on VLTI/MIDI angular diameter	\\
diam\_cohen	& mas	&  Fitted angular diameter from \citet{cohen99} \\
e\_diam\_cohen	& mas	&  Error on Cohen's angular diameter	\\
diam\_gaia	& mas	&  Estimate of angular diameter from Gaia DR2$^{a}$\\
LDD\_meas &	mas	&  Measured LDD angular diameter from JMDC	\\
e\_diam\_meas	& mas	&  Error on measured angular diameter from JMDC	\\
UDD\_meas	& mas	&  Measured UD angular diameter from JMDC	\\
band\_meas	&   &  Text describing the wavelength or band of the angular diameter measurement from JMDC \\
LDD\_est	& mas	&  Estimated LDD angular diameter from JSDC\_V2	\\
e\_diam\_est	& mas	&  Error on estimated angular diameterfrom JSDC\_V2	\\
UDDL\_est	& mas	&  Estimated UD angular diameter in$ L$ from JSDC\_V2\\
UDDM\_est	 & mas	&  Estimated UD angular diameter in $M$ from JSDC\_V2	\\
UDDN\_est	& mas	&  Estimated UD angular diameter in $N$ from JSDC\_V2	\\
Jmag &	mag	&  2MASS $J$ magnitude (1.25~$\mu$m)	\\
Hmag	& mag	& 2MASS $H$ magnitude (1.65~$\mu$m)	\\
Kmag	& mag	& 2MASS $Ks$ magnitude (2.17~$\mu$m)	\\
W4mag	& mag	&  AllWISE W4 magnitude (22.1~$\mu$m)	\\
CalFlag	&   & [0/7] Confidence flag for using this star as a calibrator in Opt. Interf. experiments from JSDC\_V2 \\
IRflag	&   & 	[0/7] Confidence flag indicating the probable presence of MIR features (excess, extent, variability)\\
nb\_Lflux	&   & Number of flux values reported in $L$	\\
med\_Lflux	& Jy	&  Median flux value in $L$	\\
disp\_Lflux	& Jy	&  Dispersion of flux values in $L$	\\
nb\_Mflux	&   & Number of flux values reported in $M$	\\
med\_Mflux	& Jy	&  Median flux value in $M$ 	\\
disp\_Mflux	& Jy	&  Dispersion of flux values in $M$	\\
nb\_Nflux	&   & Number of flux values reported in $N$	\\
med\_Nflux	& Jy	&  Median flux value in $N$	\\
disp\_Nflux	 &Jy	&  Dispersion of flux values in $N$	\\
Lcorflux\_30	& Jy	&  Uniform-disk correlated flux in $L$ for 30-m baselength	\\
Lcorflux\_100	& Jy	&  Uniform-disk correlated flux in $L$ for 100-m baselength\\
Lcorflux\_130	& Jy	&  Uniform-disk correlated flux in $L$ for 130-m baselength	\\
Mcorflux\_30	& Jy	&  Uniform-disk correlated flux in $M$ for 30-m baselength	\\
Mcorflux\_100	& Jy	&  Uniform-disk correlated flux in $M$ for 100-m baselength	\\
Mcorflux\_130	& Jy	&  Uniform-disk correlated flux in $M$ for 130-m baselength\\
Ncorflux\_30	& Jy	&  Uniform-disk correlated flux in $N$ for 30-m baselength	\\
Ncorflux\_100	& Jy	&  Uniform-disk correlated flux in $N$ for 100-m baselength\\
Ncorflux\_130	& Jy	&  Uniform-disk correlated flux in $N$ for 130-m baselength\\
\hline
\end{tabular}
\\
\begin{flushleft}
{
$^{a}${The values of teff\_gaia and diam\_gaia were determined only from the three broad-band photometric measurements used with \textit{Gaia}. The strong degeneracy between $T_{\rm eff}$ and extinction/reddening when using the broad-band photometry necessitates strong assumptions in order to estimate their values \citep[see e.g.][]{casagrande18}. One should thus be very careful in using these astrophysical parameters and refer to the papers and online documentation for guidance.}
}
\end{flushleft}
\end{table*}

 Table~\ref{table:columns} describes the columns reported in the catalogue.


\bsp	
\label{lastpage}
\end{document}